\documentclass[11pt,dvips]{article}

\usepackage{jheppub}

\usepackage{slashbox,latexsym,amssymb,amsfonts,amsmath,slashed}
\usepackage{graphicx}
\newcommand{\f}[2]{\frac{#1}{#2}}
\newcommand{\tf}[2]{{\textstyle\f{#1}{#2}}}
\newcommand{\la}{\langle}

\newcommand{\ra}{\rangle}

\newcommand{\slaD}{{\slashed D}}

\newcommand{\nyp}[3]{{\bf #1} (#3) #2}
\newcommand{\JHEP}{Jour.\ High Energy Phys.\ }
\renewcommand{\Re}{{\rm Re}\,}

\newcommand{\tr}{{\rm tr}\,}
\newcommand{\diag}{{\rm diag}\,}
\newcommand{\tdU}{{U}^{\rm (td)}}
\newcommand{\tilU}{\tilde{U}}

\title{An Anderson-like model of the QCD chiral transition}

\author[a,b]{Matteo Giordano}
\author[c]{\!\!, Tam\'as G.\ Kov\'acs}%
\author[a,b]{\!\!, and Ferenc Pittler}%

\affiliation[a]{
Institute for Theoretical Physics, E\"otv\"os University,\\
P\'azm\'any P. s\'et\'any 1/A, H-1117 Budapest, Hungary}

\affiliation[b]{
  MTA-ELTE Lattice Gauge Theory Research Group,\\ P\'azm\'any
  P. s\'et\'any 1/A, H-1117 Budapest, Hungary}

\affiliation[c]{
Institute for Nuclear Research of the Hungarian Academy of Sciences, \\
Bem t\'er 18/c, H-4026 Debrecen, Hungary 
}%

\emailAdd{giordano@bodri.elte.hu}
\emailAdd{kgt@atomki.mta.hu} 
\emailAdd{pittler@bodri.elte.hu}

\abstract{We study the problems of chiral symmetry breaking and
eigenmode localisation in finite-temperature QCD by looking at the
lattice Dirac operator as a random Hamiltonian. We recast the
staggered Dirac operator into an unconventional three-dimensional
Anderson Hamiltonian (``Dirac-Anderson Hamiltonian'') carrying
internal degrees of freedom, with disorder provided by the
fluctuations of the gauge links. 
In this framework, we identify the features relevant to chiral
symmetry restoration and localisation of the low-lying Dirac
eigenmodes in the ordering of the local Polyakov lines, and in the
related correlation between spatial links across time slices, thus
tying the two phenomena to the deconfinement transition. 
We then build a toy model based on QCD and on the Dirac-Anderson
approach, replacing the Polyakov lines with spin variables and
simplifying the dynamics of the spatial gauge links, but preserving
the above-mentioned relevant dynamical features. Our toy model
successfully reproduces the main features of the QCD spectrum and of
the Dirac eigenmodes concerning chiral symmetry breaking and
localisation, both in the ordered (deconfined) and disordered
(confined) phases. Moreover, it allows us to study separately the
roles played in the two phenomena by the diagonal and the off-diagonal
terms of the Dirac-Anderson Hamiltonian. 
Our results support our expectation that chiral symmetry restoration
and localisation of the low modes are closely related, and that both
are triggered by the deconfinement transition.}


\begin{document}

\maketitle

\section{Introduction}

The low end of the spectrum of the Euclidean Dirac operator plays an
important role in determining the properties of hadronic matter in
Quantum Chromodynamics (QCD). In particular, the spectral density
around the origin is closely tied to the fate of chiral symmetry, and
entirely determines it in the chiral limit~\cite{BC}. It is thus not
surprising that the low end of the Dirac spectrum behaves differently
in the broken and in the restored phase. The most important difference
is obviously that while in the broken phase eigenvalues accumulate
around the origin, in the restored phase the spectral density vanishes
there. Besides this, or perhaps as a consequence, the low-lying
eigenmodes display different localisation properties and statistical
behaviour. 

Numerical simulations of QCD on the lattice have shown that the
low-lying Dirac eigenmodes, while delocalised on the entire lattice
volume at low temperature~\cite{VWrev,deF}, become spatially localised
at high
temperature~\cite{GGO,GGO2,KGT,KP,BKS,KP2,feri,crit,Cossu:2014aua,Cossu:2016scb}, 
above the chiral crossover~\cite{Aoki:2005vt,Borsanyi:2010cj}.
Although most of the studies of this phenomenon have been carried out
using the staggered discretisation of the Dirac
operator~\cite{GGO,GGO2,KP,BKS,KP2,feri,crit},  
there is also evidence in simulations with overlap~\cite{KGT,BKS} and
domain wall~\cite{Cossu:2014aua,Cossu:2016scb} fermions. Let us summarise the
current knowledge about it (see Ref.~\cite{Giordano:2014qna} for a
review), focussing on the case of the staggered operator. In this case
the eigenvalues $i\lambda$ are purely imaginary and the spectrum is
symmetric with respect to zero, so it suffices to discuss $\lambda \ge
0$. Above the QCD crossover temperature, $T_c$, the low-lying quark
eigenmodes are spatially localised on the scale of the inverse
temperature, for eigenvalues below a critical point in the spectrum,
$\lambda_c$. Eigenmodes above $\lambda_c$, on the other 
hand, extend throughout the whole system. The position of the critical
point depends on the temperature, $\lambda_c=\lambda_c(T)$, and it
extrapolates to $\lambda_c=0$ at a temperature compatible with
$T_c$, as determined from thermodynamic
observables~\cite{Aoki:2005vt,Borsanyi:2010cj}. This strongly suggests 
that the appearance of localised modes is related to the
deconfinement/chiral transition. Further evidence in this respect
comes from simulations of quenched QCD with $N_c=3$~\cite{GGO2}
and $N_c=2$~\cite{KP} colours, and from the study of unimproved
staggered fermions for temporal extension $N_T=4$ in lattice
units~\cite{unimproved} reported  
in Ref.~\cite{GKKP}: both models display a genuine
deconfinement/chiral phase transition, accompanied by the appearance
of localised modes. 

As we mentioned above, the different localisation properties of the
low-lying modes in the low and high temperature phases of QCD come
along with different statistical properties. At low temperature there
is clear evidence that the eigenvalues obey the Wigner--Dyson
statistics predicted by Random Matrix Theory (RMT)~\cite{VWrev}. At
high temperature, on the other hand, the localised modes at the low
end of the spectrum fluctuate independently, thus obeying Poisson
statistics, while the delocalised eigenmodes higher up in the
spectrum, above $\lambda_c$, obey again the predictions of
RMT~\cite{KP,BKS,KP2,feri,crit}. This allows one to study the
localisation/delocalisation transition by looking at the spectral
statistics. 

In Ref.~\cite{crit} the transition in the spectrum from localised to
delocalised modes was shown to be a true second-order phase
transition, analogous to the metal-insulator transition in the
Anderson model~\cite{Anderson58,LR,EM}, which describes
non-interacting electrons in a disordered crystal. By studying the
corresponding transition in the statistical properties of the
spectrum, the correlation-length critical exponent was found to 
be compatible with that of the 3D unitary Anderson
model~\cite{nu_unitary}. ``Unitary'' refers here to the symmetry class 
of the model in the RMT classification of random matrix
ensembles~\cite{Mehta}, and the staggered Dirac operator also belongs
to this class~\cite{VWrev}. This result for the critical exponent
has been recently confirmed by a study of the multifractal properties
of the eigenmodes at criticality~\cite{UGPKV}. In the same paper it
was also found that the multifractal exponents of the critical
eigenmodes in QCD are compatible with those of the 3D unitary Anderson 
model~\cite{UV}, which further supports that the delocalisation
transitions in the two models belong to the same universality class. 

The result about the universality class of the delocalisation
transition in the high-tempe\-ra\-ture Dirac spectrum of QCD is quite
surprising. As a statistical model, lattice QCD at finite temperature
is four-dimensional (although anisotropic), while the 3D unitary
Anderson model is, well, three-dimensional. Moreover, the Hamiltonian
of the 3D unitary Anderson model consists of a diagonal, on-site
random potential, and of nearest-neighbour hopping terms with fixed
and uniform strength (and random phases). The staggered Dirac
operator, on the other hand, consists entirely of off-diagonal
nearest-neighbour hopping terms, depending on the gauge transporter on
the corresponding lattice link, and it is known from the condensed-matter
literature that off-diagonal noise is much less effective in inducing
localisation, with a large amount of disorder required to this
end~\cite{offdiag,offdiag2,GGC}. 

A possible explanation of the fact that the two models share the same
universality class has been proposed in Ref.~\cite{GKP}, elaborating
on a previous proposal made in Ref.~\cite{BKS} for the mechanism
leading to localisation. The main idea is that high-temperature QCD
effectively contains a source of 3D on-site disorder, which consists
of the phases of the local Polyakov line at a given spatial
point. Above $T_c$ the Polyakov lines get ordered along the identity, 
with local fluctuations away from it. The Polyakov line phases affect
the quark eigenmodes through effective boundary conditions in the
temporal direction, and make favourable for the quark wave function to
``live'' on the ``islands'' of unordered Polyakov lines. As a random
matrix model, the Dirac operator in high-temperature QCD is thus
effectively 3D with diagonal noise. Support to the viability of this
mechanism was obtained in Ref.~\cite{BKS} by studying the correlation
of the Dirac eigenfunctions with the fluctuations of the Polyakov loop 
on $SU(2)$ gauge configurations. In Ref.~\cite{GKP} we looked for a
different kind of evidence: we constructed, and studied numerically, a
QCD-inspired toy model which should display localisation precisely
through the proposed mechanism, but in a much simplified setting. This
``Ising-Anderson'' model is essentially obtained by removing the
temporal direction, thus working in three-dimensional space (i.e., a
single time slice), and by mimicking the effect of the Polyakov lines
on the quark wave functions with a (continuous) spin model of the
Ising class in the ordered phase, used to generate the appropriate
diagonal noise. This model adequately reproduces the main features of
localisation, in particular their qualitative dependence on the amount
of ``islands'' of ``wrong'' spins in the ``sea'' of ordered spins, as
expected from the proposed explanation.  

Although the Ising-Anderson model of Ref.~\cite{GKP} provides a
satisfactory qualitative description of localisation in the
high-temperature phase of QCD, it fails completely at describing the
low-temperature phase. Indeed, simulations in the disordered phase of
the underlying Ising model (not reported in Ref.~\cite{GKP}, but see
Section \ref{sec:num} below for similar results) fail to 
reproduce the most distinctive feature of the Dirac spectrum at low
temperature, namely the presence of a nonzero spectral density around
the origin, which leads to the spontaneous breaking of chiral
symmetry~\cite{BC}. Instead, a sharp gap separates the lowest
eigenvalues from the origin, both when the underlying spin model is in
the ordered phase, and when it is in the disordered phase. While in
the former case this is in rough qualitative agreement with the small
density of near-zero modes in QCD at high temperature, in the latter
case this is at odds with the existing numerical results. On the other
hand, the mechanism through which the Polyakov lines affect the quark
wave functions, i.e., the effective boundary conditions, is in
principle at work at all temperatures, and only the presence of order
in the (relevant) Polyakov-line configurations, or lack thereof,
distinguish the two phases. As we have already said above, the {\it
  effectiveness} of the  ``sea/islands'' mechanism devised in
Refs.~\cite{BKS,GKP} can explain the {\it presence} of localised modes
at high temperature. For the explanation to be complete, the {\it
  ineffectiveness} of this same mechanism should also explain the {\it
  absence} of localised modes at low temperatures. This requirement is
even more compelling in light of the apparently very close relation
between localisation and the QCD crossover: the reason for the
ineffectiveness of the mechanism at low temperature is likely to be
also the reason for the finiteness of the spectral density at the
origin.  

Above we said that the effective boundary conditions are in principle
at work at all temperatures but, as a matter of fact, they turn out to
be irrelevant at low temperatures, where QCD looks ``effectively
4D''. The ineffectiveness of the boundary conditions at low
temperatures clearly entails the ineffectiveness of the
``sea/islands'' mechanism for localisation, and so it may seem
hopeless to try to give a simultaneous description of the QCD
crossover and of the appearance of localised modes by means of a 3D
model. Still, one can technically treat the quark degrees of freedom
along the compactified temporal direction as internal degrees of
freedom of a quark living in 3D space. In doing this, one can in
principle make the temporal extent of the lattice as small as possible
in lattice units, thus reducing the amount of such internal degrees of
freedom to a minimum. In this setting, whether the system is
effectively 3D or 4D becomes a question about the correlation among
the internal degrees of freedom. For this question to make any sense
at all, the number of such degrees of freedom has clearly to be at
least two. Perhaps a more natural way of phrasing this is in terms of
the temporal correlation length, and the minimal number of time slices
for which one can ask if they are strongly correlated is obviously two. 

With this (in hindsight rather obvious) insight, our purpose in this
paper is to build a refined toy model, aimed at describing the
(supposedly) simultaneous appearance of localised modes and recovery
of chiral symmetry, as signalled by the vanishing of the spectral
density at the origin.\footnote{Since chiral symmetry is explicitly
  broken by the quark masses, by ``recovery'' we mean here the 
  dramatic change in the strength of the symmetry breaking, which,
  loosely speaking, corresponds to the disappearance of the effects
  that can be ascribed to spontaneous symmetry breaking.} The
motivation is twofold. On the one hand, we want to implement the 
``sea/islands'' mechanism in a model that reproduces QCD more
faithfully (at the qualitative level), in order to make the
case of Refs.~\cite{BKS,GKP} stronger. On the other hand, we want to
investigate the connection between localisation and the
deconfinement/chiral transition in a simple and controllable
setting. This could also lead to some insight into the chiral
transition, and into its relation to deconfinement, from the point of
view of the QCD Dirac operator as a random matrix model, independently
of the issue of localisation. Indeed, since localisation of the low
modes and restoration of chiral symmetry take place together, and this
happens near the deconfinement transition, it is very likely that they
are both triggered by the ordering of the gauge configurations, so
that some mechanism could be devised which would explain both
phenomena. 

Here is the plan of the paper. Before building the toy model, in
Section~\ref{sec:SF} we cast the staggered Dirac operator into a
three-dimensional Hamiltonian, with internal degrees of freedom
corresponding to colour and to the lattice temporal momenta. In this
way the connection with Anderson-type models is made transparent. In
Section~\ref{sec:toy} we write down explicitly our toy model, which we
study numerically in Section~\ref{sec:num}. Finally, in
Section~\ref{sec:concl} we discuss our results, state our conclusions
and comment about future prospects. Some technical details are
discussed in Appendix \ref{sec:app}. 

\section{Staggered Dirac operator as a random matrix model}
\label{sec:SF}

In this Section we recast the Dirac operator in lattice QCD, $\slaD$,
as the Hamiltonian of an unconventional three-dimensional Anderson
model. We work here with the staggered Dirac operator for simplicity. 

The basic idea is to split the ``Hamiltonian'' $H= -i\slaD$ into a
``free'' and an ``interaction'' part, $H=H_0+H_I$, and then work in
the basis of the ``unperturbed'' eigenvectors of $H_0$. At finite
temperature the temporal direction is compactified and therefore
singled out, and so the physically most sensible choice is to identify
the free Hamiltonian with the temporal hoppings. This leaves the
spatial hoppings as the (spatially isotropic) interaction part. 
We thus define
\begin{equation}
  \label{eq:split}
  \begin{aligned}
  H_0: &~\text{temporal hoppings,} &&&  (H_0)_{xx'} &= 
  \f{\eta_4(\vec x)}{2i}\left\{U_4(t,\vec 
    x)\delta_{t+1,t'} 
      - U_{-4}(t,\vec x)\delta_{t-1,t'}\right\}\delta_{\vec x\vec
      x{}'}\,, \\
 H_I: &\text{~spatial hoppings,} &&&     (H_I)_{xx'} &= \sum_{j=1}^3 
  \f{\eta_j(\vec x)}{2i}\left\{U_j(t,\vec x)\delta_{\vec x+\hat\jmath,\vec x{}'}
- U_{-j}(t,x)\delta_{\vec x-\hat\jmath,\vec
  x{}'}\right\}\delta_{tt'}\,,
  \end{aligned}
\end{equation}
where $x=(t,\vec x)$ with $t$ the Euclidean ``time'', $0\le t\le
N_T-1$, and the staggered phases $\eta_\mu(\vec x)$ are given by 
\begin{equation}
  \label{eq:etas}
\eta_\mu(\vec x)=(-1)^{ \left(\sum_{\nu=1}^\mu x_\nu\right) -x_\mu }\,,
\end{equation}
and depend only on the spatial coordinates, $0\le x_i\le L-1$,
$i=1,2,3$. Both $N_T$ and $L$ must be even. 
We take the gauge links $U_{\mu}(t,\vec x)\in SU(N_c)$ for generality,
and denote the backward links by $U_{-4}(t,\vec x)\equiv
U^\dag_{4}(t-1,\vec x)$ and $U_{-j}(t,\vec x)\equiv U^\dag_{j}(t,\vec
x-\hat \jmath)$, $j=1,2,3$. 
Periodic boundary conditions are understood for the gauge links, while
on the quark wavefunctions antiperiodic boundary conditions in the
temporal direction, and periodic boundary conditions in the spatial
directions, are imposed. In Eq.~\eqref{eq:split} colour indices are
suppressed. 

The eigenvectors of $H_0$ are easily determined. To this end, let us
define the following gauge transporters in the temporal direction,
\begin{equation}
  \label{eq:tdep_P}
  P_0(\vec x) \equiv \mathbf{1}\,,\qquad P_{t+1}(\vec x)\equiv P_{t}(\vec x)
  U_4(t,\vec x)\,,~ 0\le t\le N_T-1\,,
\end{equation}
with $P_{N_T}(\vec x)=P(\vec x)$ the usual (untraced) Polyakov line
starting at $t=0$ and winding around the temporal direction. Let
furthermore $\varphi_a(\vec x)$ be the (ortho)normalised eigenvectors
of $P(\vec x)$,  
\begin{equation}
  \label{eq:6}
    P(\vec x) \varphi_a(\vec x) = e^{i\phi_a(\vec x)} \varphi_a(\vec
    x)\,,\qquad
    \varphi_a^\dag(\vec x)\varphi_b(\vec x) = \delta_{ab}\,.
\end{equation}
Here $\vec x$ runs over the whole 3-volume, $a=1,\ldots,N_c$
with $N_c$ the number of colours, and each $\varphi_a$ has $N_c$ colour
components, $(\varphi_a)_i$. The eigenvalues $e^{i\phi_a(\vec x)}$
have unit absolute value and satisfy $\prod_{a}e^{i\phi_a(\vec x)}
=1$. The eigenvectors of the Polyakov line can be used to build 
the eigenvectors $\psi_0^{\vec x\,a\,k}$ of $H_0$. These are localised
on a single spatial point, $\vec x$, have a well-defined temporal
momentum, $k$, and carry a colour quantum number, $a$, and read (in 
the coordinate basis)
\begin{equation}
  \label{eq:5}
  \psi_0^{\vec x\,a\,k}(t,\vec y) = \f{1}{\sqrt{N_T}}\,  \delta_{\vec
    x\vec y}\, e^{i\omega_{ak}(\vec x)t}
\,P_t^\dag(\vec x)\varphi_a(\vec x)
\,, 
\end{equation}
with colour- and space-dependent effective Matsubara frequencies,
$\omega_{ak}(\vec x)$, 
\begin{equation}
  \label{eq:mats_freq_x}
      \omega_{ak}(\vec x) = {\textstyle\f{1}{N_T}}(
      \pi + \phi_a(\vec x)
        + 2\pi k)\,,\qquad k=0,1,\ldots,N_T-1\,.
\end{equation}
The form of $\omega_{ak}(\vec x)$ results from imposing temporal
antiperiodic boundary conditions on the fermions, and from the
presence of nontrivial Polyakov lines which modify the free-field
result. The $\psi_0^{\vec x\,a\,k}$ are eigenvectors of $H_0$ with
``unperturbed'' eigenvalues $\lambda_0^{\vec x\,a\,k}$ given by
\begin{equation}
  \label{eq:7}
\lambda_0^{\vec x\,a\,k}=\eta_4(\vec x) \sin \omega_{ak}(\vec
  x)\,.
\end{equation}
Notice that
\begin{equation}
  \label{eq:7bis}
  \sin \omega_{a k'}(\vec x) = -\sin \omega_{ak}(\vec
  x) ~~\text{if}~~ k'=\textstyle\left(\f{N_T}{2}+k\right)_{N_T} \,,
\end{equation}
where we have denoted $\left(a+b\right)_{N_T} \equiv a+b\mod N_T$.
In the basis $\{\psi_0^{\vec x\,a\,k}\}$, the operator $H_I$ has
vanishing diagonal elements, and only nearest-neighbour hopping terms 
(i.e., connecting eigenvectors localised on nearest-neighbour sites),
\begin{equation}
  \label{eq:int_ham}
  \begin{aligned}
    \protect{[H_I]}_{\vec x\,a\,k,\vec y\,b\,l} &=
\sum_{j=1}^3 \delta_{\vec x+\hat \jmath,\vec y}[H_I]_{\vec x\,a\,k,\vec x + \hat
  \jmath\,b\,l}  + 
\delta_{\vec x-\hat \jmath,\vec y}[H_I]_{\vec  x\,a\,k,\vec x - \hat 
  \jmath\,b\,l} \,,\\
[H_I]_{\vec x\,a\,k,\vec x\pm \hat \jmath\,b\,l} &=
\pm \f{\eta_j(\vec x)}{2i} \f{1}{N_T} \sum_{t=0}^{N_T-1}
e^{i[\omega_{bl}(\vec x\pm \hat \jmath)-\omega_{ak}(\vec x)]t} 
  \left[ \tdU_{\pm j}(t,\vec x)\right]_{ab}\,,
  \end{aligned}
\end{equation}
where
\begin{equation}
  \label{eq:tdiag_gauge}
  \left[\tdU_{\pm j}(t,\vec x)\right]_{ab} =\varphi_a^\dag(\vec x)
   P_t(\vec x) U_{\pm j}(t,\vec x)
   P_t^\dag(\vec x\pm \hat \jmath) \varphi_b(\vec x\pm \hat \jmath)\,.
\end{equation}
Notice that $\tdU_{\pm j}(t,\vec x)$ is just the usual link
variable in the temporal diagonal gauge, i.e., the temporal gauge
where the Polyakov lines have been diagonalised.\footnote{It is
  understood here that the eigenvectors of the Polyakov lines are
  chosen so that $\det_{ai} (\varphi_a)_i = 1~\forall \vec x$ (or at
  least constant over space). Indeed, for a general choice one can
  only prove that $|\det \tdU_{\pm j}| = 1$. Notice that choosing
  another basis of eigenvectors (namely, changing phases and order of
  the eigenvectors) corresponds to a unitary transformation of the
  Hamiltonian, which therefore does not affect the spectrum.} 

The full Hamiltonian in the basis of ``unperturbed'' eigenvectors, 
$\{\psi_0^{\vec x\,a\,k}\}$, will be denoted by ${\cal H}$, and it
carries space, colour, and temporal-momentum indices, $[{\cal H}(\vec 
x,\vec y)]_{ak,bl}$. Suppressing the latter two indices, we will write
\begin{equation}
  \label{eq:H_new_b}
  \begin{aligned}
    {\cal H}(\vec x,\vec y)&= 
    {\cal D}(\vec x,\vec y) +
\sum_{j=1}^3\f{\eta_j(\vec x)}{2i}\left\{{\cal V}_{+j}(\vec x,\vec
  y)-{\cal V}_{-j}(\vec x,\vec y)\right\}
\\
&=    \delta_{\vec x, \vec y} D(\vec x)
    + \sum_{j=1}^3\f{\eta_j(\vec x)}{2i}\left\{
      \delta_{\vec x+\hat\jmath,\vec y} V_{+j}(\vec x)
      -\delta_{\vec x-\hat\jmath,\vec y} V_{-j}(\vec x)
\right\}\,,
  \end{aligned}
\end{equation}
where we have introduced the following notation,
\begin{equation}
  \label{eq:H_new_b_elem}
  \begin{aligned}
    \left[D(\vec x)\right]_{ak,bl} &= 
\lambda_0^{\vec x\,a\,k}\delta_{ab}\delta_{kl}=
\eta_4(\vec
    x)\sin\omega_{ak}(\vec x)\delta_{ab}\delta_{kl}\,,\\
    \left[V_{\pm j}(\vec x)\right]_{ak,bl} &=
\f{1}{N_T} \sum_{t=0}^{N_T-1}
e^{i\f{2\pi t}{N_T}(l-k)} \left[\tilU_{\pm j}(t,\vec
  x)\right]_{ab}\,,\\ 
\left[\tilU_{\pm j}(t,\vec x)\right]_{ab} &=
e^{i\f{t}{N_T}[\phi_{b}(\vec x\pm \hat \jmath)-\phi_{a}(\vec
  x)]} 
\left[\tdU_{\pm j}(t,\vec x)\right]_{ab}\,.
  \end{aligned}
\end{equation}
Hermiticity implies ${\cal V}^\dag_{\pm j}={\cal
  V}^{\phantom{\dag}}_{\mp j}$, as one can also verify
explicitly. Moreover, $\tilU_{\pm j}(t,\vec x)$ are unitary matrices
in colour space, and $V_{\pm j}(\vec x)$ are unitary matrices in
(joint) colour and temporal-momentum space. The phases $\phi_a(\vec
x)$ are defined only modulo $2\pi$, so that in order to fully specify
${\cal H}$ one needs to pick a convention. The simplest and most
sensible possibility is to take $\phi_a(\vec x)\in [-\pi,\pi)$ for
$a=1,\ldots,N_c-1$, and impose $2\pi q(\vec x)\equiv \sum_a
\phi_a(\vec x)=0$ for all $\vec x$. This is what we do from now on,
unless otherwise specified. In this way one avoids introducing
spurious non-uniformities in the Hamiltonian that might obscure the
important features. In principle, however, any other choice, possibly
different on different lattice sites, is legitimate. One can show that 
the Hamiltonians obtained with different choices are related by
unitary transformations, so that the spectrum does not depend on the 
convention used, as it should be. If one and the same convention is
adopted for $q(\vec x)\in \mathbb{Z}$ on all lattice sites, then
$\tilU_{\pm j}(t,\vec x)$ and $V_{\pm j}(\vec x)$ are also
unimodular. More details on these issues are given in Appendix 
\ref{sec:app}.

The Hamiltonian Eq.~\eqref{eq:H_new_b} looks like that of a 3D
Anderson model, but with an antisymmetric rather than symmetric
hopping term, and moreover carrying internal degrees of freedom. 
We will sometimes refer to it as the Dirac-Anderson Hamiltonian.
Due to  the fluctuations of the gauge links from configuration to
configuration, both diagonal and off-diagonal disorder are
present. The amount of disorder is thus controlled by the size 
of such fluctuations, and thus ultimately by the temperature of the
system (and by the lattice spacing). 

The diagonal disorder originates entirely from the Polyakov lines. In
contrast with the usual Anderson Hamiltonian, changing the amount of
disorder does not change the strength of the diagonal term, since the 
``unperturbed'' eigenvalues are always bounded by one in absolute
value. On the other hand, on the two sides of the deconfinement
transition the shape of their distribution is different, with an
enhancement of ``unperturbed'' eigenvalues corresponding to the
trivial phase at high temperature. Moreover, in the high-temperature
phase there is long-range order in the diagonal term, as a consequence
of the long-range order in the Polyakov-line configuration.

The off-diagonal disorder in the hopping terms is mostly determined by 
the spatial links. As with the diagonal disorder, while the overall
``size'' of the hopping term does not change with temperature, as it
is a unitary matrix in any case, its typical matrix structure changes
considerably across the transition. Indeed, the most interesting
property of the hopping term is that when $\tilU_{\pm j}(t,\vec x)
\equiv \tilU_{\pm j}(\vec x)$ is time-independent, then $V_{\pm
  j}(\vec x)$ is block-diagonal in temporal-momentum space, i.e., 
\begin{equation}
  \label{eq:V_corr}
        \left[V_{\pm j}(\vec x)\right]_{ak,bl} =
\left[\tilU_{\pm j}(\vec x)\right]_{ab}
 \f{1}{N_T} \sum_{t=0}^{N_T-1}
e^{i\f{2\pi t}{N_T}(l-k)} = \left[\tilU_{\pm j}(\vec x)\right]_{ab}
\delta_{kl}\,.
\end{equation}
For this to happen we need both $\phi_{b}(\vec x\pm \hat
\jmath)-\phi_{a}(\vec x)=0~\forall a,b$ and $\tdU_{\pm j}(t,\vec
x)=\tdU_{\pm j}(0,\vec x)~\forall t$. This is the case when the
neighbouring Polyakov lines are both close to the
identity,\footnote{The special role of the identity among the center 
  elements is a consequence of our convention $q(\vec x)=0$ for the
  phases, which implies that the first condition is equivalent to
  $\phi_{a}(\vec x)=\phi_{b}(\vec x\pm \hat\jmath)=0~\forall a,b$. A
  different convention $q(\vec x)=n$ would pick up another center
  element, and since this simply corresponds to a change of basis, the
  following statements remain true for any choice of the
  high-temperature vacuum, provided that the appropriate basis is
  used.}  
which in turn causes a strong (local) correlation among spatial link
variables. At low temperatures the Polyakov lines are disordered, and
so this does not occur often: we thus expect that typically $V_{\pm
  j}(\vec x)$ will have non-negligible off-diagonal terms in
temporal-momentum space, which leads to strong mixing of the
wave-function components corresponding to different temporal momenta. 
At high temperature, on the other hand, the Polyakov lines get ordered,
and there are large spatial regions where this is approximately true:
in these regions $V_{\pm j}(\vec x)$ gets ordered along the identity in
temporal-momentum space, and so in these regions the components of the
wave function corresponding to different temporal momenta are coupled
only weakly.\footnote{In contrast to this, when the spatial links are
  perfectly anticorrelated, i.e., $\tilU_{\pm j}(t,\vec x) =
  (-1)^t\tilU_{\pm j}(\vec x)$, one finds 
  \begin{equation}
    \label{eq:V_anticorr}
    \left[V_{\pm j}(\vec x)\right]_{ak,bl} =
    \left[\tilU_{\pm j}(\vec x)\right]_{ab}
    \f{1}{N_T} \sum_{t=0}^{N_T-1}
    e^{i\f{2\pi t}{N_T}(l-k - \f{N_T}{2})} = \left[\tilU_{\pm j}(\vec
      x)\right]_{ab} 
    \delta_{\left(k+\f{N_T}{2}\right)_{N_T}l}\,,
  \end{equation}
  i.e., a given temporal-momentum component $k$ mixes only with the
  ``opposite'' component $\left(k+\f{N_T}{2}\right)_{N_T}$ [see
  Eq.~\eqref{eq:7bis}].} In other words, at high temperature we expect
the ``correlation length'' in temporal-momentum space to become
shorter than the size of the system (again, in temporal-momentum
space): this is what typically happens when a transition to a 
disordered phase takes place. Keep in mind that here the system under
consideration are the quark eigenfunctions, and not the gauge fields,
and moreover that this shortening of the ``correlation length'' is a
local effect (although taking place in the whole ``sea'' of ordered
Polyakov lines).

Finally, we want to remark that the spectrum of ${\cal H}$ is
obviously symmetric with respect to $\lambda=0$. In the new basis,
this is seen as a consequence of the anticommutation relation 
$\{{\cal Q},{\cal H}\}=0$ of the Hamiltonian with a certain unitary
matrix ${\cal Q}$. Moreover, in the case of gauge group $SU(2)$ the
Hamiltonian is invariant under an antiunitary transformation ${\cal
  T}$ with ${\cal T}^2=-1$, which implies a double (Kramers)
degeneracy of the spectrum, and moreover that the Hamiltonian belongs
to the symplectic class of random Hamiltonians~\cite{Mehta}. These are 
nothing but the analogues of the properties of the usual staggered
Dirac operator, only expressed in a different basis. Details on the
form of ${\cal Q}$ and ${\cal T}$ in the new basis are summarised in
Appendix \ref{sec:app}.

\section{Construction of the toy model}
\label{sec:toy}

In this Section we want to build explicitly a simple model describing
the change both in the localisation properties and in the spectral
density of the low Dirac modes, starting from the Dirac-Anderson
Hamiltonian, i.e., the reformulation of the Dirac operator described
in the previous Section.

\subsection{Brief review of gauge dynamics}
\label{sec:gdrev}

We begin by reviewing the features of the dynamics of gauge theories
that we expect to be relevant for the two above-mentioned phenomena.
As we have already pointed out in the Introduction, the Polyakov line
is expected to play an important role in the localisation of low modes
at high temperature. As is well known, in pure gauge theory the center
symmetry of the action is spontaneously broken in the deconfined phase
at high temperature, where the Polyakov lines get ordered along one of
the center elements. Fluctuations away from the ordered value form
``islands'' within a ``sea'' of ordered Polyakov lines. In the low
temperature, confined phase the center symmetry is restored and
Polyakov lines are disordered. Dynamical fermions in the fundamental
representation break the center symmetry explicitly but softly, so
that they do not change this picture, besides selecting the trivial
vacuum in the high-temperature phase. 

As we observed at the end of the previous Section, the ordering of the
Polyakov lines in the high-temperature phase of QCD will induce strong
correlations between spatial links at the same spatial point but on
neighbouring time slices. This is a local effect, i.e., it depends on
the ordering of the Polyakov lines at the spatial points connected by
the given link. To see this, it is convenient to work in the temporal
diagonal gauge [see Eq.~\eqref{eq:tdiag_gauge}], where the
contributions to the Wilson gauge action read (up to a constant factor) 
\begin{equation}
  \label{eq:sp_link_WA}
  \Re\tr \left\{\tdU_{j}(t,\vec x)\left[ \tdU_{j}{}^\dag(t+1,\vec x)
+  \tdU_{j}{}^\dag(t-1,\vec x)\right]\right\} + {\rm SP}\,,
\qquad t\neq 0,N_T-1\,,
\end{equation}
away from the temporal boundary, while at the temporal boundary we
have 
\begin{equation}
  \label{eq:sp_link_WA_2}
  \begin{aligned}
    &  \Re\tr\left\{\tdU_{j}(N_T-1,\vec x)\left[ P(\vec x+\hat \jmath)
    \tdU_{j}{}^\dag(0,\vec x) P^\dag(\vec x) 
+   \tdU_{j}{}^\dag(N_T-2,\vec x)\right]\right\} +
{\rm SP}\,,\\ 
&  \Re\tr\left\{ \tdU_{j}(0,\vec x)\left[ \tdU_{j}{}^\dag(1,\vec x)
+  P^\dag(\vec x+\hat \jmath)
\tdU_{j}{}^\dag(N_T-1,\vec x) P(\vec x)\right]\right\} + {\rm SP}\,.  
  \end{aligned}
\end{equation}
Here ${\rm SP}$ stands for the contribution of the spatial plaquettes.  
Although only $\tdU_{j}(0,\vec x)$ and $\tdU_{j}(N_T-1,\vec x)$
interact directly with the Polyakov lines, this effect propagates to
the spatial links on the other time slices, as they are coupled
according to Eq.~\eqref{eq:sp_link_WA}. Moreover,
Eq.~\eqref{eq:sp_link_WA_2} shows that the dynamics of the Polyakov
lines is affected by the backreaction of the spatial links. 

\subsection{From QCD to the toy model}
\label{sec:qcdtotoy}

Let us now turn to the explicit construction of the model. 
We want to study the spectrum of a random Hamiltonian of the form 
\begin{equation}
  \label{eq:toy_Ham}
  {\cal H}^{\rm toy}(\vec x,\vec y) =  d(\vec
x)\delta_{\vec x \vec y} + \sum_{j=1}^3
  \f{\eta_j(\vec x)}{2i}\left(
v_{+ j}(\vec x)\delta_{\vec x+\hat\jmath,\vec y}
- v_{- j}(\vec
x)\delta_{\vec x-\hat\jmath,\vec y}
\right)\,,
\end{equation}
with $d(\vec x)$ and $v_{\pm j}(\vec x)$ respectively diagonal and
unitary in joint colour and temporal-momen\-tum space. This is of the
same form as the staggered Dirac operator in the basis of
``unperturbed'' eigenvectors, Eqs.~\eqref{eq:H_new_b} and
\eqref{eq:H_new_b_elem}, and both the diagonal and the hopping terms
will be modelled on Eq.~\eqref{eq:H_new_b_elem}, replacing the
Polyakov line phases $\phi_a(\vec x)$ and the spatial links
$\tdU_j(t,\vec x)$ with similar quantities in the toy model. This
implies in particular that the toy model Hamiltonian will have a 
symmetric spectrum, and will belong to the same class of random
Hamiltonians as the one found in ($N_c$-colour) QCD.  

Since our purpose is to build a model simpler than QCD, yet displaying 
the same behaviour concerning the phenomena of localisation of
eigenmodes and accumulation of eigenvalues near the origin, we will
take QCD as a starting point and eliminate all those features that we
deem irrelevant. We will first of all neglect the backreaction of the
quark eigenmodes in the partition function, omitting the fermion
determinant (i.e., making the {\it quenched} approximation), since it
is known that these phenomena are present in pure gauge theory as
well. 

Next, since we aim at reproducing these phenomena only qualitatively,  
we will simplify the dynamics of the toy model analogues of
$\phi_a(\vec x)$ and $\tdU_j(t,\vec x)$ with respect to QCD. As in
Ref.~\cite{GKP}, the main simplifying idea is to mimic the effect of
the Polyakov line phases on the quark wave functions by spin-like
variables. However, in this work we want to achieve a closer
resemblance to the actual dynamics of the phases. To this end, we want
to design the spin model so that the effective potential for the
magnetisation, in the ordered phase, is similar to that for the 
Polyakov line phases in QCD~\cite{Yaffe:1982qf,DeGrand:1983fk}, or
more generally in an $SU(N_c)$ gauge theory. The potential should
therefore develop $N_c$ minima in the ordered phase, corresponding to 
the $N_c$ Polyakov-loop vacua. A possibility is to choose $N_c$
complex spin variables $s_{\vec x}$, corresponding to the $N_c$ 
eigenvalues of an $SU(N_c)$ Polyakov line, that satisfy
\begin{equation}
  \label{eq:phase_def}
  |s_{\vec x}^a| = 1\,,~~a=1,\ldots,N_c\,,\qquad
\prod_{a=1}^{N_c} s_{\vec x}^a=1\,,
\end{equation}
and which obey the dynamics determined by the following
Hamiltonian,
\begin{equation}
  \label{eq:sp_mod_ham}
  \begin{aligned}
\beta H_{\rm noise} & = \beta \sum_{\vec x,j,a}
    \left[1-{\f{1}{N_c}}\Re\tr\{
    p(x+\hat\jmath)^\dag p(x)\}\right]+{\f{hN_c}{N_c-1}}
    \sum_{\vec x}\left(1-\bigg|{\f{1}{N_c}} \tr
      p(x)\bigg|^2\right)\\
 &= \f{\beta}{2N_c} \sum_{\vec x,j,a} |s^a_{\vec x
  + \hat\jmath}-s^a_{\vec      x}|^2  +\f{hN_c}{N_c-1}
\sum_{\vec x}\left(1-\bigg|\f{1}{N_c} \sum_a s^a_{\vec
     x}\bigg|^2\right)\,,
  \end{aligned}
\end{equation}
where $p(x) = \diag(s_{\vec x}^1,\ldots,s_{\vec x}^{N_c})$.
The $N_c$-dependent factors are chosen for convenience. The first term
corresponds to a lattice sigma-model possessing a global
$[U(1)]^{N_c-1}$ symmetry. The second term mimics the absolute value
squared of the trace of the Polyakov line (i.e., the Polyakov loop),
and at $h\neq 0$ breaks the symmetry down to ${\mathbb
  Z}_{N_c}$.\footnote{Further unbroken symmetries are the one under  
  ``charge conjugation'', $s^a_{\vec x}\to s^a_{\vec x}{}^*$, and the
  one under permutations of the $s^a_{\vec x}$ with respect to $a$.}
This residual 
symmetry can hold dynamically or be spontaneously broken, with
precisely $N_c$ vacua $s_{\vec x}^a=e^{i\f{2\pi}{N_c}k}~\forall \vec
x,a$, with $k=0,\ldots,N_c-1$.\footnote{The minimum of $H_{\rm noise}$
  is achieved for spatially uniform phases (modulo $2\pi$), $s^a_{\vec 
    x}=e^{i\phi^a}$, satisfying $\phi^a=\phi^b \mod 2\pi~\forall
  a,b$. For a general parameterisation of the phases the constraint
  reads $\sum_a \phi^a = 0 \mod 2\pi$, which leads to
  $\phi^a=\f{2\pi}{N_c}k \mod 2\pi~\forall a$, $k=0,\ldots,N_c-1$.} 
Parameterising the spins as follows,
\begin{equation}
  \label{eq:phase_def_2}
s_{\vec x}^a = e^{i\phi_{\vec x}^a}\,,\quad  \phi_{\vec x}^a\in 
[
-\pi,\pi
)
\,,~~a=1,\ldots,N_c-1\,,\qquad
\sum_{a=1}^{N_c} \phi_{\vec x}^a=0\,,
\end{equation}
the Hamiltonian can be recast as
\begin{equation}
  \label{eq:sp_mod_ham_2}
  \beta H_{\rm noise} =  
  -\f{\beta}{N_c} \sum_{\vec x,j,a}  \cos(\phi^a_{\vec
    x + \hat\jmath}-\phi^a_{\vec x})
  - \f{2h}{N_c(N_c-1)}\sum_{\vec x,a<b}\cos(\phi^a_{\vec
    x}-\phi^b_{\vec x}) \,, 
\end{equation}
up to an irrelevant additive constant. The dynamics of the phases
$\phi_{\vec x}^a$ resembles qualitatively that of the Polyakov line
phases: while at low $\beta$ the (complex) magnetisations
$m^a=L^{-3}\sum_{\vec x}s^a_{\vec x}$ vanish on average, for large
enough $\beta$ the system transitions to an ordered phase, with
$\phi_{\vec x}^a$ aligning to one of the vacuum values discussed
above. Small and large $\beta$ in the spin model thus correspond to
small and large temperatures in QCD. It is then natural to take the
diagonal term $d(\vec x)$ to simply be $D(\vec x)$ with the Polyakov
line phases replaced by $\phi_{\vec x}^a$, i.e., 
\begin{equation}
  \label{eq:toy_Ham_2}
  \left[d(\vec x)\right]_{ak,bl}  =  \eta_4(\vec
  x)\sin\tf{\pi + \phi_{\vec x}^a + 2\pi k}{N_T} \delta_{ab}\delta_{kl}
  \,.
\end{equation}
Notice that $\phi_{\vec x}^a$ will obey their own independent dynamics,
unaffected by the analogues of $\tdU_j(t,\vec x)$. 

Let us now turn to the hopping terms. These are defined by replacing
Polyakov line phases and gauge links in Eq.~\eqref{eq:H_new_b_elem}
with the phases $\phi_{\vec x}^a$ and with appropriate $SU(N_c)$
matrices $u_j(t,\vec x)$, respectively:
\begin{equation}
  \label{eq:toy_Ham_3}
      \left[v_{\pm j}(\vec x)\right]_{ak,bl} =
\f{1}{N_T} \sum_{t=0}^{N_T-1}
e^{i\f{2\pi t}{N_T}(l-k)} 
e^{i\f{t}{N_T}[\phi^{b}_{\vec x\pm \hat \jmath}-\phi^{a}_{\vec
  x}]} 
\left[u_{\pm j}(t,\vec x)\right]_{ab}\,,
\end{equation}
with $u_{-j}(t,\vec x)=u_{j}^\dag(t,\vec x-\hat\jmath)$. The last step
is to define the dynamics of $u_j(t,\vec x)$. The important feature we
want to mimic from QCD are the local correlations between spatial
links across time slices induced by the Polyakov lines. On the other
hand, we expect the correlations induced by the spatial plaquettes in
Eqs.~\eqref{eq:sp_link_WA} and \eqref{eq:sp_link_WA_2} to be less
important. We will thus drop them from the action, keeping only the
contributions of the temporal plaquettes. The Boltzmann weight for the
configurations of the spatial links thus factorises, with each factor
involving only the spatial links at a given spatial point and along a
given spatial direction. Explicitly, the dynamics of the toy model
links $u_j(t,\vec x)$ will be governed by the following action, 
\begin{equation}
  \label{eq:toy_Ham_4_0}
  S_u = -\f{1}{2}\sum_{\vec x}\sum_{t=0}^{N_T-1}\sum_{j=1}^3 \delta
  S_j(t,\vec x)\,, 
\end{equation}
where
\begin{equation}
  \label{eq:toy_Ham_4}
  \begin{aligned}
    \delta S_j(t,\vec x)&= \hat\beta\,  \Re\tr \left\{u_{j}(t,\vec x) 
      \left[ u_{j}^\dag(t+1,\vec x) +  u_{j}^\dag(t-1,\vec
        x)\right]\right\}\,,  \quad t\neq 0,N_T-1\,, \\
    \delta S_j(N_T-1,\vec x)&=\hat\beta\,\Re\tr \left\{
      u_{j}(N_T-1,\vec x) \left[ p(\vec x+\hat \jmath) 
        u_{j}^\dag(0,\vec x) p^\dag(\vec x) + u_{j}^\dag(N_T-2,\vec x)
      \right]\right\} \,,\\
    \delta S_j(0,\vec x) &= \hat\beta\,
    \Re\tr \left\{u_{j}(0,\vec x) \left[u_{j}^\dag(1,\vec x) +
        p^\dag(\vec x+\hat \jmath) 
        u_{j}^\dag(N_T-1,\vec x) p(\vec x)
      \right]\right\}\,, 
  \end{aligned}
\end{equation}
where $\hat\beta$ is a constant playing the role of the gauge
coupling. Expectation values are defined as follows: 
\begin{equation}
  \label{eq:toy_Ham_5}
  \la {\cal O} \ra = \f{\int D\phi \,e^{-\beta H_{\rm noise}[\phi]} \left[\f{\int
        Du \, e^{-S_u[\phi,u]}{\cal O}[\phi,u]}{\int Du\,
        e^{-S_u[\phi,u]}}\right]}{\int D\phi \,e^{-\beta H_{\rm noise}[\phi]}} \,,
\end{equation}
where $\int D\phi=\prod_{\vec x,a} \int_{-\pi}^{+\pi}d\phi^a_{\vec x}$ and $Du
=\prod_{\vec x,t,j} du_j(t,\vec x)$, with $du_j(t,\vec x)$ the Haar
measure. Notice that the average over $u_j(t,\vec x)$ is done at fixed
$\phi^a_{\vec x}$, i.e., there is no backreaction of the link
variables on the spins, and $\phi^a_{\vec
  x}$ acts as a background field for $u_j(t,\vec x)$.

\subsection{Minimal toy model}
\label{sec:mintoy}

Let us describe the model in detail in the simplest case, namely
taking $N_T=2$ and $N_c=2$, i.e., the minimal possible values. This is
the model we have employed in the numerical study discussed in the
next Section. 

For $N_c=2$, the basic variables are the complex spins $s_{\vec
  x}\equiv s_{\vec x}^1 = s_{\vec x}^{2*}=e^{i\phi_{\vec x}}$, with
$\phi_{\vec x}\equiv \phi^1_{\vec x}=-\phi^2_{\vec x}$, and the
$SU(2)$ link variables $u_{j}(t,\vec x)$, $t=0,1$. The noise
Hamiltonian governing the spin dynamics, Eq.~\eqref{eq:sp_mod_ham_2},
simplifies to
\begin{equation}
  \label{eq:sp_mod_ham_NC2}
  \beta H_{\rm noise}^{N_c=2} =  
  -\beta \sum_{\vec x,j}  \cos(\phi_{\vec x + \hat\jmath}-\phi_{\vec x})
  - h\sum_{\vec x}\cos(2\phi_{\vec x}) \,. 
\end{equation}
The $U(1)$ symmetry of the $XY$ model, represented by the first term,
is broken to $\mathbb{Z}_2$ by a nonzero ``external field'', $h$,
appearing in the second term. We thus expect our spin model
to belong to the universality class of the three-dimensional Ising
model. 

Concerning the dynamics of the link variables, for $N_T=2$ there is a 
single term $\delta S_{j}(0,\vec x)= \delta S_{j}(1,\vec x) \equiv
\delta S_{j}(\vec x)$ determining the Boltzmann weight for $j$-links
at $\vec x$, namely 
\begin{equation}
  \label{eq:sp_link_WA_3}
  \begin{aligned}
 \delta S_{j}(\vec x)&=  \hat\beta\,\Re\tr \left\{u_{j}(1,\vec x) 
   \left[u_{j}(0,\vec x) 
     +   p(\vec x) u_{j}(0,\vec x)p^\dag(\vec x+\hat \jmath)
   \right]^\dag\right\}\\
 &=  \hat\beta\,\Re\tr \left\{u_{j}(0,\vec x)
   \left[u_{j}(1,\vec x) 
     +   p^\dag(\vec x) u_{j}(1,\vec x)p(\vec x+\hat \jmath)
   \right]^\dag\right\} \,, 
  \end{aligned}
\end{equation}
where $p(\vec x)=\diag(e^{i\phi_{\vec x}},e^{-i\phi_{\vec x}})$, and
the action for the links reads simply $S_u=-\sum_{\vec x,j}\delta
S_{j}(\vec x)$. Averages are defined according to
Eq.~\eqref{eq:toy_Ham_5} above. It is worth reminding the reader that
while in QCD there is a single coupling that enters the dynamics of
both the Polyakov lines and the spatial links, in the toy model
$\beta$ and $\hat\beta$ can be varied independently.

The toy model Hamiltonian ${\cal H}^{\rm toy}(\vec x,\vec y)$,
Eq.~\eqref{eq:toy_Ham}, mimicking the QCD Dirac operator, consists 
of a diagonal and a hopping term, both containing disorder. For the
on-site, diagonal noise terms $d(\vec x)$, since 
\begin{equation}
  \label{eq:sin_NT2}
  \sin\f{\pi \pm \phi_{\vec x} + 2\pi k}{2} = (-1)^k\cos\f{\phi_{\vec
      x}}{2} \,, \quad k=0,1\,,
\end{equation}
we have simply
\begin{equation}
  \label{eq:onsite_2_2_exp}
  d(\vec x) =
  \eta_4(\vec x)\cos\f{\phi_{\vec x}}{2}\begin{pmatrix}
    \mathbf{1} & \mathbf{0}
    \\ \mathbf{0} &- \mathbf{1}
    \end{pmatrix}\,.
\end{equation}
As for the hopping terms $v_{j}(\vec x)$, they read explicitly
\begin{equation}
  \label{eq:hopping_2_2_exp}
  \begin{aligned}
    v_{j}(\vec x) &=
    \begin{pmatrix}
      v^{(+)}_{j}(\vec x)  & v^{(-)}_{j}(\vec x) \\
      v^{(-)}_{j}(\vec x)  & v^{(+)}_{j}(\vec x)  
    \end{pmatrix}\,,\\
    v^{(\pm)}_{j}(\vec x)  &=  
    \f{1}{2}\begin{pmatrix}
      [ u_{j}(0,\vec x)]_{11}  
      \pm e^{\f{i \Delta_{j}(\vec x)}{2}}    [ u_{j}(1,\vec x)]_{11} &
      [ u_{j}(0,\vec x)]_{12}  \pm e^{-\f{i\Sigma_{j}(\vec x)}{2}}
      [ u_{j}(1,\vec x)]_{12} \\
      [ u_{j}(0,\vec x)]_{21}  \pm e^{\f{i\Sigma_{j}(\vec x)}{2}}
      [ u_{j}(1,\vec x)]_{21} &
      [ u_{j}(0,\vec x)]_{22}  
      \pm e^{-\f{i \Delta_{j}(\vec x)}{2}}    [ u_{j}(1,\vec x)]_{22}
    \end{pmatrix}\,,
  \end{aligned}
\end{equation}
where
\begin{equation}
  \label{eq:sp_link_WA_6}
  \Delta_{j}(\vec x) = \phi_{\vec x +\hat\jmath} -
  \phi_{\vec x}\,,
  \qquad \Sigma_{j}(\vec x) = \phi_{\vec x +\hat\jmath} + \phi_{\vec x}\,.
\end{equation}
Similarly, $v_{-j}(\vec x)$ are defined by replacing $j\to -j$ in
Eqs.~\eqref{eq:hopping_2_2_exp} and \eqref{eq:sp_link_WA_6}. 
In Eqs.~\eqref{eq:onsite_2_2_exp} and \eqref{eq:hopping_2_2_exp} the
matrix blocks correspond to the temporal-momentum indices $k=0,1$
(row-wise) and $l=0,1$ (column-wise)  of Eqs.~\eqref{eq:toy_Ham_2} and
\eqref{eq:toy_Ham_3}.

\begin{figure}[t]
  \centering
  \includegraphics[width=0.7\textwidth]{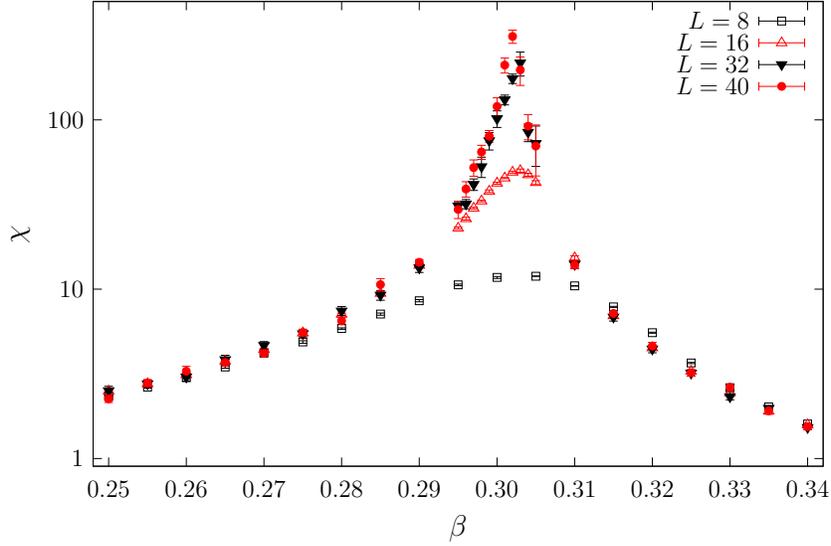} 
  \caption{Magnetic susceptibility of the spin model. Here $h=1.0$.}
  \label{fig:1}
\end{figure}

\section{Numerical results}
\label{sec:num}

In this Section we discuss our numerical results for the toy model
defined in the previous Section, both in the ordered and in the
disordered phases of the underlying spin model. For simplicity, we
have studied the ``minimal'' case $N_c=N_T=2$. The ``gauge coupling'',
$\hat \beta$, was fixed to $\hat \beta=5.0$. Since we are interested
mostly in the dependence on $\beta$, we have fixed the coefficient $h$
of the symmetry-breaking term to $h=1.0$.

\begin{figure}[t]
  \centering
  \includegraphics[width=0.7\textwidth]{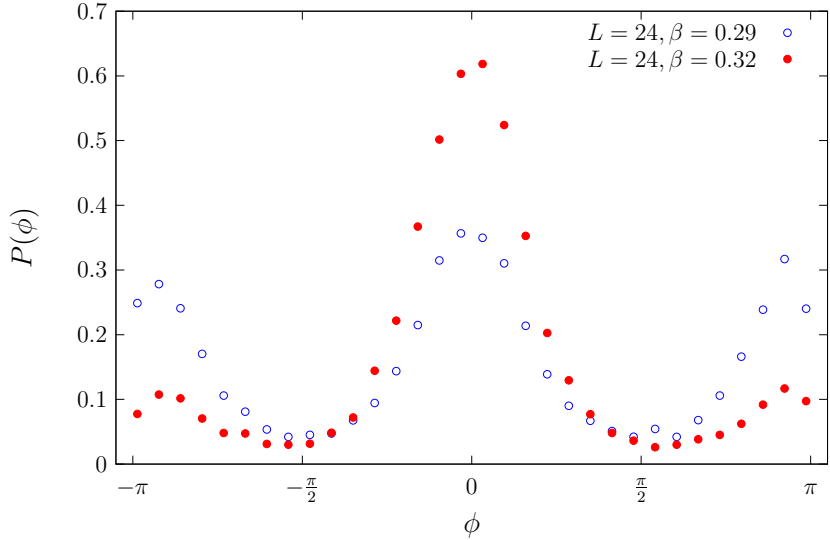} 
  \caption{Distribution of spin phases in a single typical configuration of
    the spin system in the disordered (circles) and in the ordered
    (full points) phases. Here $h=1.0$ and $L=24$.}
  \label{fig:sd}
\end{figure}

We have first studied the spin model on its own to determine the
corresponding phase structure. In Fig.~\ref{fig:1} we show the
magnetic susceptibility $\chi$ of the spin model,
\begin{equation}
  \label{eq:mag_susc}
\chi =  L^{-3}\left[\left\la({\textstyle\sum_x}\Re s_x)^2\right\ra -
  (\left\la{\textstyle\sum_x}\Re s_x\right\ra)^2\right] 
=  L^{-3}\left[\left\la({\textstyle\sum_x}\cos \phi_x)^2\right\ra -
  (\left\la{\textstyle\sum_x}\cos \phi_x\right\ra)^2\right]\,,
\end{equation}
as a function  of $\beta$. A phase transition is expected to occur in
the thermodynamic limit at a critical $\beta_c$, with $\beta_c\approx
0.3$. For $\beta\lesssim 0.29$ one is safely in the disordered phase,
while for $\beta\gtrsim 0.31$ one is in the ordered phase, and
finite-size effects should not affect the qualitative behaviour of our
random toy Hamiltonian ${\cal H}^{\rm toy}$. In Fig.~\ref{fig:sd} we
show the distribution of phases in a single typical configuration
below and above the transition. The tendency of the system to get
ordered is evident. 

\begin{figure}[t]
  \centering
  \includegraphics[width=0.7\textwidth]{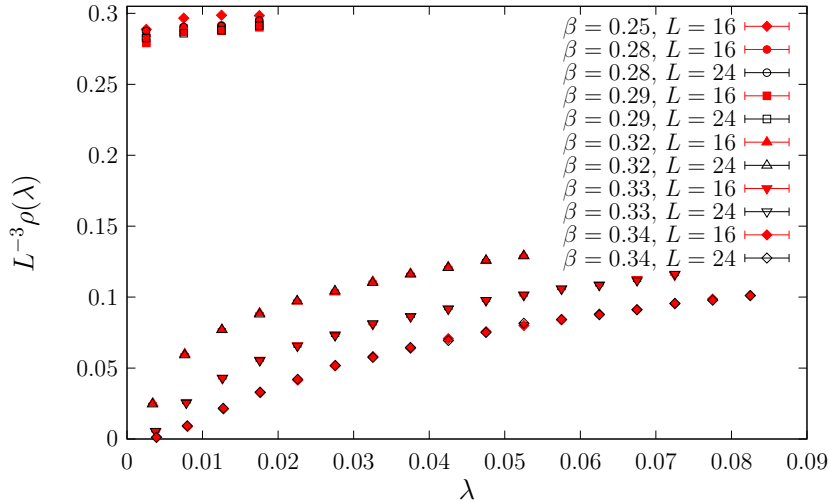} 
  \caption{Spectral density of the toy Hamiltonian for
    several values of $\beta$ and different volumes. Here $h=1.0$ and
    $\hat \beta = 5.0$.}
  \label{fig:2}
\end{figure}

Since our toy model is quenched, in the ordered phase of the spin
model we have to select the appropriate vacuum by hand. The
appropriate vacuum is of course that in which the phase of the 
magnetisation $m=N_c^{-1}L^{-3}\sum_{\vec x,a}s^a_{\vec x} = |m|e^{i\varphi_m}$
is zero, corresponding to the trivial Polyakov loop sector selected by
fermions in QCD. This is done in practice only at the level of the
random Hamiltonian, by ``rotating'' the spins, i.e., aligning their
phase to zero, when the spin model is in a different vacuum. In the
case $N_c=2$ considered in this paper this is easy to implement. For
typical configurations in the ordered phase, the (complex)
magnetisation is usually close to being real, i.e.,
$e^{i\varphi_m}\simeq \pm 1$. When $\cos\varphi_m<0$, we ``rotate''
all the spins by replacing $s_{\vec x} \to -s_{\vec x}$. In terms of
the phases 
$\phi_{\vec x}\in [-\pi,
\pi)$, this is implemented through
\begin{equation}
  \label{eq:phase_rotation}
  \phi_{\vec x} \to 
  \phi_{\vec x} - \pi\,{\rm sign}(\phi_{\vec x})\,, \qquad {\rm
    sign}(0)\equiv 1\,.
\end{equation}
The spectral density $\rho(\lambda)$ of the random Hamiltonian in the
two phases is shown in Fig.~\ref{fig:2}: while in the disordered phase
($\beta<\beta_c$) there is an accumulation of eigenvalues near the
origin, so that (presumably) $L^{-3}\rho(0)\neq 0$ in the
thermodynamic limit, in the ordered phase ($\beta>\beta_c$) this
region is depleted, and $\rho(0)=0$. The dependence on $\beta$ is
rather mild in the disordered phase, while the spectral density gets
rapidly suppressed as $\beta$ grows in the ordered phase. The expected
connection between magnetisation in the spin system and ``chiral''
transition in the spectrum of the random Hamiltonian indeed shows up.

\begin{figure}[t]
  \centering
  \includegraphics[width=0.7\textwidth]{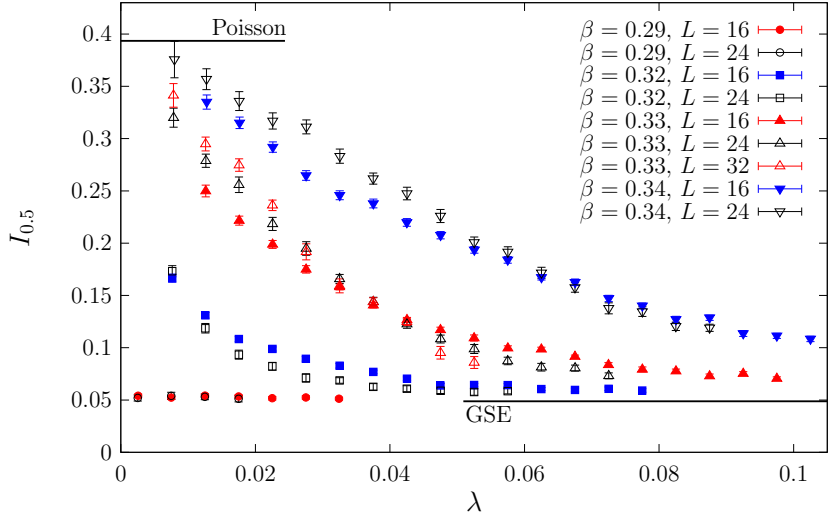} 
  \caption{Spectral statistics $I_{0.5}$ along the spectrum for
    several values of $\beta$ and different volumes. Here $h=1.0$ and
    $\hat \beta = 5.0$.} 
  \label{fig:3}
\end{figure}

In order to understand the nature of the lowest eigenmodes in the two
phases, it is convenient to study the statistical properties of the
corresponding eigenvalues. In fact, localised modes are expected to
fluctuate independently, so that the corresponding eigenvalues should
obey Poisson statistics. Delocalised modes, on the other hand, mix
strongly under fluctuations and are expected to obey the appropriate
Wigner--Dyson statistics, which in the case at hand is the one
corresponding to the Gaussian Symplectic Ensemble (GSE). A convenient
observable to distinguish the two cases is the spectral statistics 
$I_{0.5}$~\cite{SSSLS,HS}, 
\begin{equation}
  \label{eq:i05_def}
  I_{0.5} \equiv \int_0^{0.5} ds\,P_{\rm ULSD}(s)\,,
\end{equation}
where $P_{\rm ULSD}(s)$ is the probability distribution of the
unfolded level spacings $s_j = \f{\lambda_{j+1}-\lambda_j}{\la
  \lambda_{j+1}-\lambda_j\ra}$, where $\la \lambda_{j+1}-\lambda_j\ra$
is the average level spacing in the spectral region corresponding to
the given level. Unfolding is a mapping of the eigenvalues that makes
the spectral density equal to 1 throughout the spectrum. In practice,
we have ordered the eigenvalues obtained in all the explored spin
configurations and replaced them by their rank divided by the number 
of configurations.\footnote{Only one eigenvalue is kept in each
  degenerate pair.} 
 The $P_{\rm ULSD}(s)$ is known both in the case of
Poisson statistics, where it is the exponential distribution $P_{\rm
  P}(s)=\exp(-s)$, and in the case of the GSE, where it is very
precisely approximated by the symplectic Wigner surmise $P_{\rm
  GSE}$~\cite{Mehta},  
\begin{equation}
  \label{eq:symp_ws}
  P_{\rm GSE}(s) = \left(\f{64}{9\pi}\right)^3 s^4
  \exp\left(-\f{64}{9\pi}s^2\right)\,. 
\end{equation}
The quantity $I_{0.5}$ is sensitive to the behaviour of $P_{\rm
  ULSD}(s)$ near $s=0$, and so it is very different for Poisson (where 
$P_{\rm P} \sim 1$ near $s=0$) and GSE ($P_{\rm GSE}\sim s^4$)
statistics. Indeed, $I_{0.5}^{\rm GSE}\simeq 0.0487$ for the GSE,
while for the Poisson ensemble $I_{0.5}^{\rm P}\simeq 0.393$. The
choice of the upper limit of integration in Eq.~\eqref{eq:i05_def} 
is made in order to maximise the difference between these two values,
as $P_{\rm P}$ and $P_{\rm GSE}$ cross near $s=0.5$. In
Fig.~\ref{fig:3} we show the behaviour of $I_{0.5}$ as one moves along
the spectrum, i.e., computing $I_{0.5}$ locally, using only
eigenvalues in disjoint bins of fixed width, and assigning the result
to the average of the eigenvalues in that bin. While in the disordered
phase one finds Wigner--Dyson statistics throughout the whole
spectrum, in the ordered phase the lowest modes have near-Poisson
statistics, and become more independent as the volume is
increased. Above a $\beta$-dependent point $\lambda_c$ in the
spectrum, the modes have near-Wigner--Dyson behaviour, and more and
more so when the volume is increased. This hints at a
localisation/delocalisation transition in the spectrum taking place in
the thermodynamic limit. As the system is made more ordered,
$\lambda_c$ increases, which is in agreement with our expectations:
qualitatively, $\lambda_c$ should behave like the spatial average of
the lowest effective Matsubara frequency~\cite{GKP}, which indeed
grows as the ordering of the system is increased. Our toy model is
therefore able to reproduce localisation of the low modes in the
ordered phase, and the qualitative dependence of the ``mobility edge''
$\lambda_c$ on the ordering of the system.

The results discussed above show that our toy model successfully
reproduces the important features of QCD for what concerns
localisation and ``chiral symmetry'' breaking/resto\-ra\-tion, i.e., 
the accumulation or not of eigenvalues near the origin. This indicates
that we have indeed kept all the important aspects of the Dirac
operator and of the gauge dynamics in our model, which can thus be
used to gain some reliable qualitative insight into the properties of
the Dirac eigenvalues and eigenmodes. Therefore, although the
following discussion deals explicitly with the toy model, it can be
directly translated to the physically relevant case of QCD by
replacing ``spins'' with Polyakov line phases. 

\subsection{Variations on the toy model}

We want now to test how the features of the toy model that we borrowed
from QCD affect the ``chiral'' transition, i.e., the drop in the
spectral density near the origin, and localisation of the lowest modes.
To this end, we study some {\it ad hoc} modifications of the toy model.

The ``chiral'' transition and localisation of the lowest modes are
clearly connected in our toy model, as they are evidently tied to the
magnetisation of the spin system. The magnetisation affects our toy
model Hamiltonian in two ways: it creates a ``sea'' of ordered spins,
where ``islands'' of fluctuations provide an  ``energetically''
convenient place for an eigenmode to localise~\cite{BKS,GKP}; and it
locally correlates the gauge links on different time slices, thus
leading to the approximate local decoupling of different temporal-momentum
components of the eigenfunctions. It is an interesting question how
important each of these two effects is for ``chiral symmetry''
breaking and localisation.

\begin{figure}[t]
  \centering
  \includegraphics[width=0.49\textwidth]{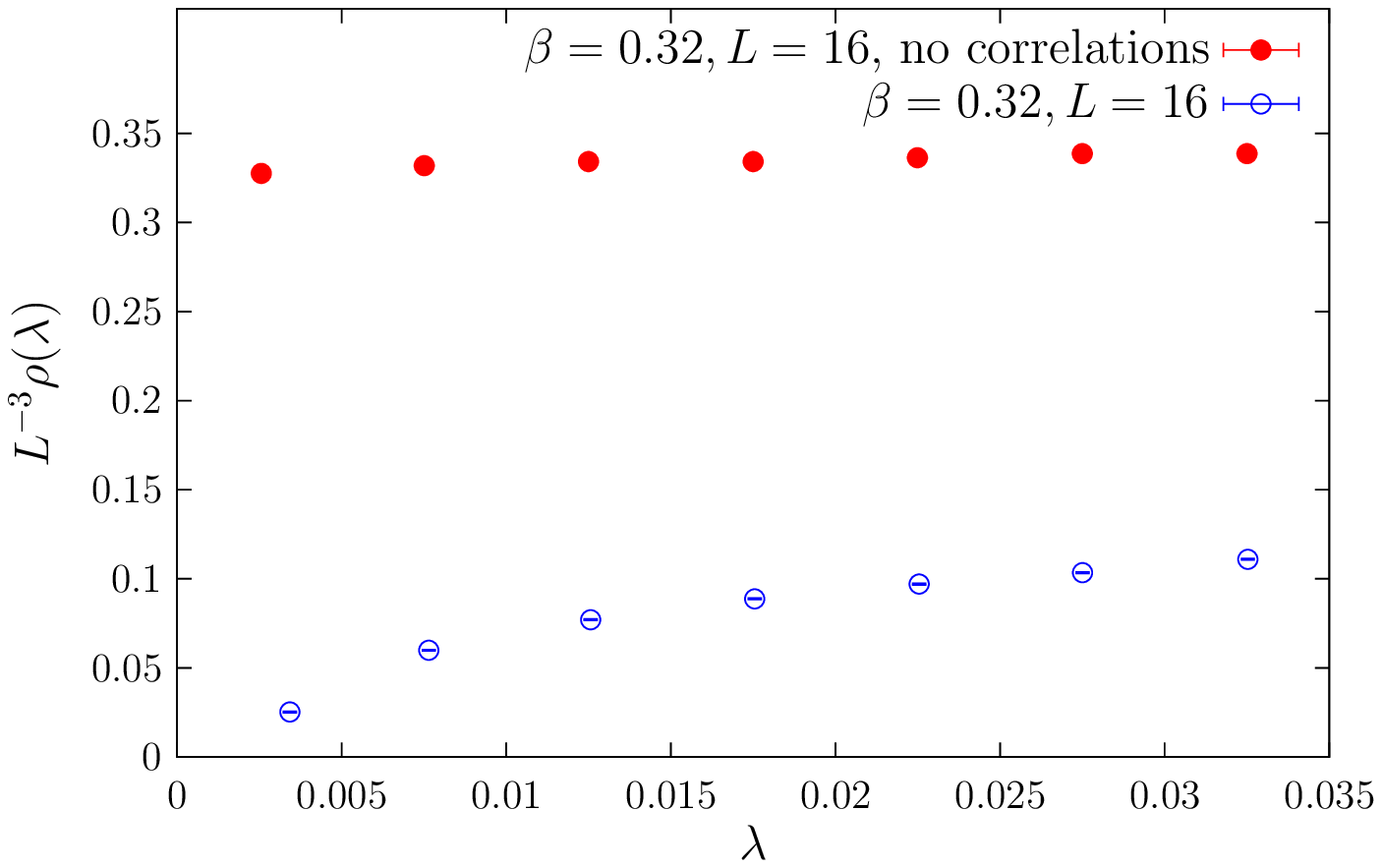} \hfil
  \includegraphics[width=0.49\textwidth]{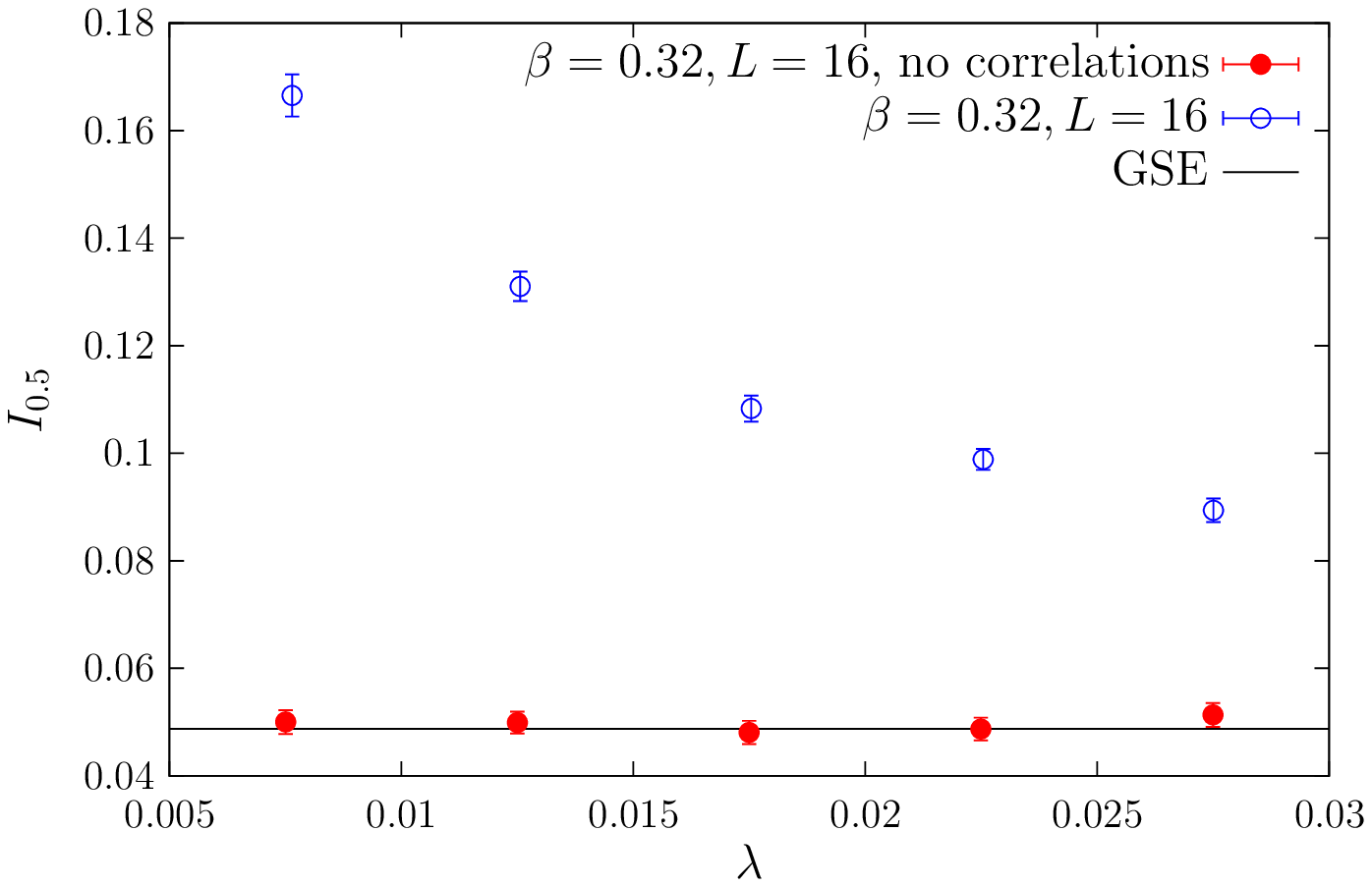} 
  \caption{Comparison between the toy model in the ordered phase with
    (circles, $\hat\beta=5.0$) and without (full points,
    $\hat\beta=0.0$) correlation between gauge 
    links on different time slices: spectral density (left panel) and
    $I_{0.5}$ (right panel). Here $L=16$ and $\beta=0.32$.}
  \label{fig:add1}
\end{figure}

As a matter of fact, the appearance of ``islands'' alone is not enough
to produce either ``chiral symmetry'' restoration or localisation. In
Fig.~\ref{fig:add1} we show the spectral density and the spectral
statistic $I_{0.5}$ obtained in the ordered phase at $\beta=0.32$ when
setting $\hat\beta=0$, i.e., imposing no correlation between the time
slices. In this case, despite the presence of ``islands'' in the spin
configurations, the spectral density at the origin remains finite, and
the low-lying eigenmodes do not localise. This means that to restore
``chiral symmetry'' one needs that the mixing of different
temporal-momentum components of the quark wave functions be suppressed
to some extent. Only then can the ``islands'' effectively act as
localising centers for the low modes, since, loosely speaking, the
lowered spectral density would make it difficult to mix for modes
localised in different regions.  

\begin{figure}[t]
  \centering
  \includegraphics[width=0.49\textwidth]{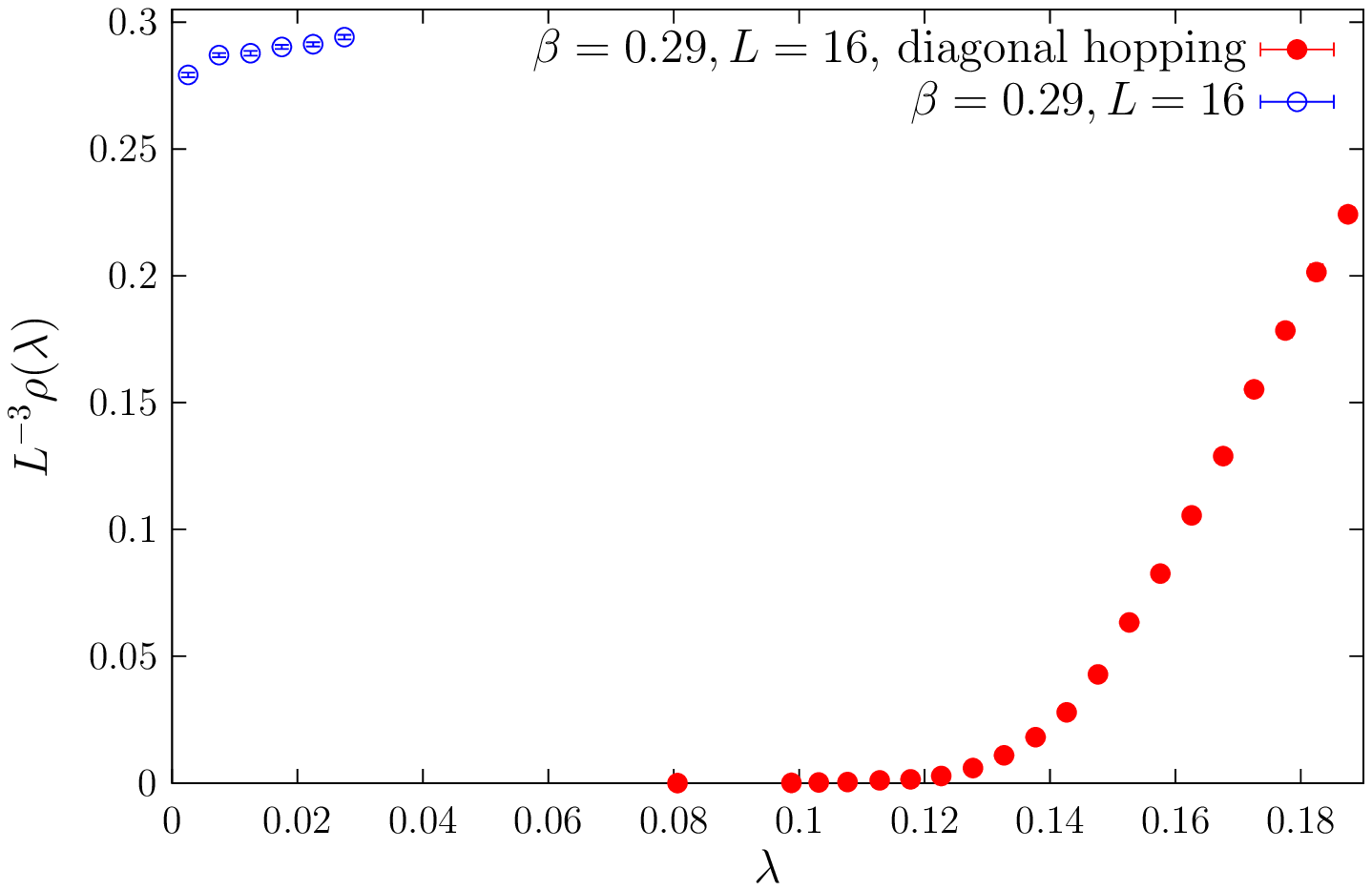} \hfil
  \includegraphics[width=0.49\textwidth]{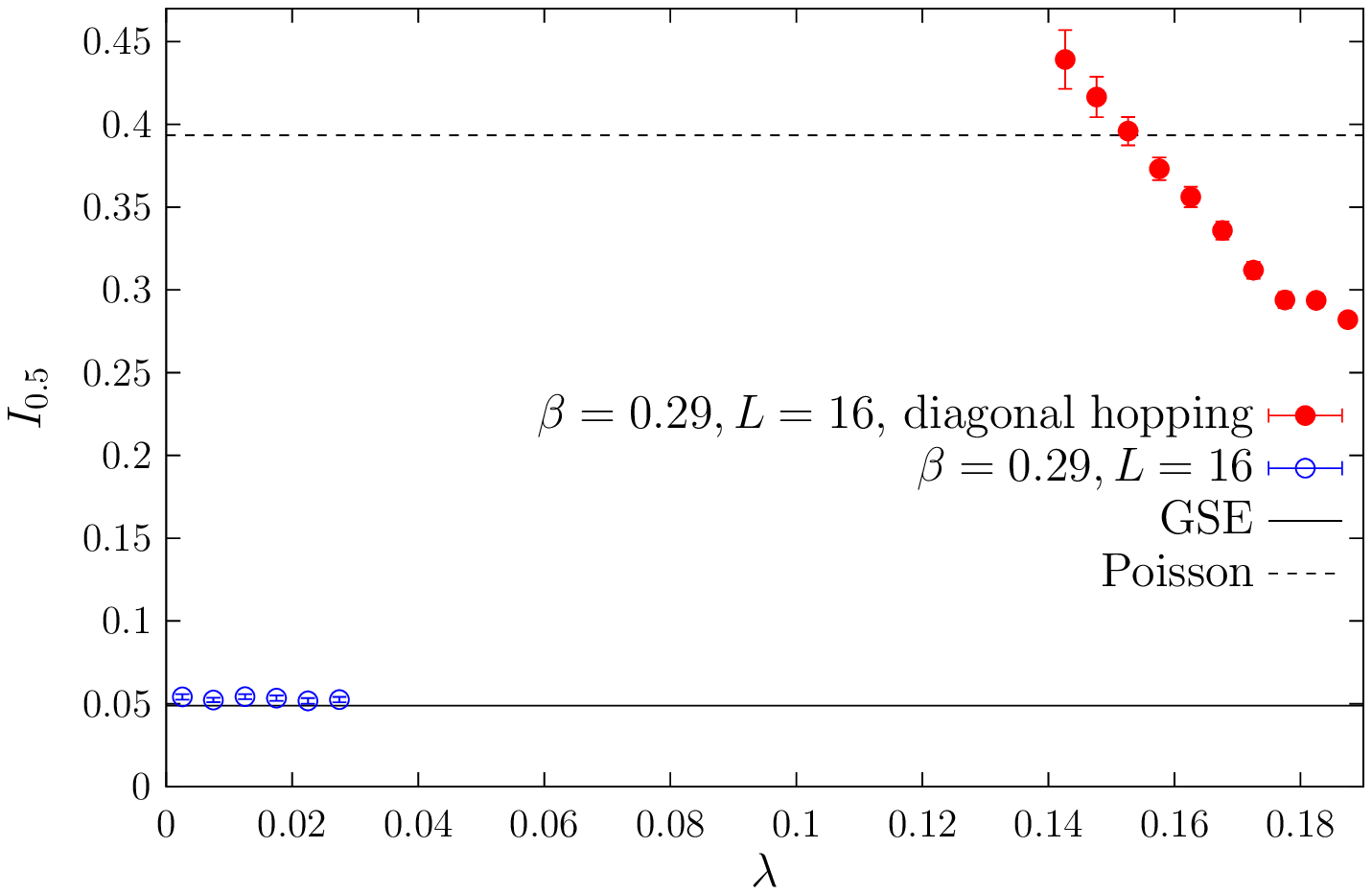} 
  \caption{Comparison between the toy model in the disordered phase with
    (circles) and without (full points) off-diagonal hopping terms
    in temporal-momentum space: spectral density (left panel) and
    $I_{0.5}$ (right panel). Here $L=16$ and $\beta=0.29$.}
  \label{fig:add2}
\end{figure}

On the other hand, in the disordered phase the absence of ``islands''
is not sufficient to ensure ``chiral symmetry'' breaking and prevent
localisation of the lowest modes. An essential ingredient for the
accumulation of eigenvalues around the origin is the fact that the
hopping term has typically sufficiently large off-diagonal components
in temporal-momentum space. In Fig.~\ref{fig:add2} we show the
spectral density obtained in the disordered phase at $\beta=0.29$ when 
setting the off-diagonal part of the hopping term to
zero.\footnote{Notice that in this case $v_{\pm j}(\vec x)$ is not a
  unitary matrix anymore. This however does not affect the important
  properties of the toy Hamiltonian, namely hermiticity, symmetry of
  the spectrum, and the antiunitary symmetry.} The Hamiltonian in this
case is block-diagonal in temporal-momentum space, with blocks ${\cal
  H}^{[\pm]}$ of the form 
\begin{equation}
  \label{eq:diag_hop_Ham}
{\cal H}^{[\pm]}(\vec x,\vec y) = 
[\pm] \eta_4(x) c(\vec x)\,\delta_{\vec x \vec y}
 + \sum_{j=1}^3
  \f{\eta_j(\vec x)}{2i}\left(
v_{+ j}^{(+)}(\vec x)\, \delta_{\vec x+ \hat\jmath, \vec y}
- v_{- j}^{(+)}(\vec x)\, \delta_{\vec x - \hat\jmath, \vec y}
\right)\,, 
\end{equation}
where $[c(\vec x)]_{ab}=\cos\tf{\phi_{\vec x}}{2} \,\delta_{ab}$, and
${v}_{\pm j}^{(+)}$ have been defined in
Eq.~\eqref{eq:hopping_2_2_exp}. Apart from the presence of colour,
and of disorder in the hopping terms,
this is precisely the Hamiltonian considered in Ref.~\cite{GKP}. In
this case the spectrum displays a sharp gap near the origin. 
In Fig.~\ref{fig:add2} we also show how the spectral statistic
$I_{0.5}$ changes along the spectrum. Our results show that the
lowest-lying modes are localised, while higher up in the spectrum
they delocalise, as signalled by the decrease of $I_{0.5}$. 
The reason why $I_{0.5}$ does not tend to the GSE value in the bulk,
but remains clearly above it, can be ascribed to the fact that the
spectra of ${\cal H}^{[\pm]}$ are separately symmetric {\it on
  average} with respect to the origin. The proof of this property can
be easily obtained extending the one reported in Ref.~\cite{GKP} to
the case of nontrivial hopping terms and in the presence of
colour. Combining the two spectra together to obtain that of ${\cal
  H}^{\rm toy}$, one expects that the latter will have an approximate
double degeneracy (on top of the Kramers degeneracy) of the
eigenvalues. This naturally leads to an enhancement of the ULSD near
the origin (levels like to lie closer than on average) with respect to
the Wigner--Dyson distribution, and therefore to an enhancement of
$I_{0.5}$.  

The results above show that the mixing of different temporal momentum
components is a necessary condition to have chiral symmetry breaking
in the disordered phase, and that its (local) attenuation is a
necessary condition to have restoration in the ordered phase. This 
entails that the ``minimal'' model studied here is indeed minimal, if
one wants to reproduce qualitatively the features of the Dirac
spectrum and of the corresponding eigenmodes. It is thus not possible
to neglect entirely the temporal direction, as in Ref.~\cite{GKP}, if
one wants to correctly describe the disordered phase. Furthermore, in
the present setting one cannot neglect the correlation of spatial
links across time slices if one wants restoration of chiral symmetry
and localisation of the low modes in the ordered phase.\footnote{In
  Ref.~\cite{GKP} this was achieved in a different setting, but the
  resulting spectral density had a sharp rather than soft gap around
  the origin.} However, one cannot certainly conclude that those
described above are also sufficient conditions, especially because of
the extreme nature of the two modifications of the toy model that we
have studied. In a realistic situation, correlations across time
slices are always present to some extent, but never perfect, so that
mixing of the temporal components of the quark wave functions is never
completely free or strictly forbidden. One thus expects some other
effect to compete against mixing, with the fate of ``chiral symmetry''
and localisation being decided by the strongest of the two
effects. The obvious candidate is the nature of the diagonal disorder,
i.e., of the distribution of the unperturbed eigenvalues: for a low
density of small unperturbed modes it should be more difficult to
accumulate eigenvalues around the origin, and conversely for a high
density of such unperturbed modes a weak mixing of temporal momentum
components could be sufficient. 

\begin{figure}[t]
  \centering
  \includegraphics[width=0.49\textwidth]{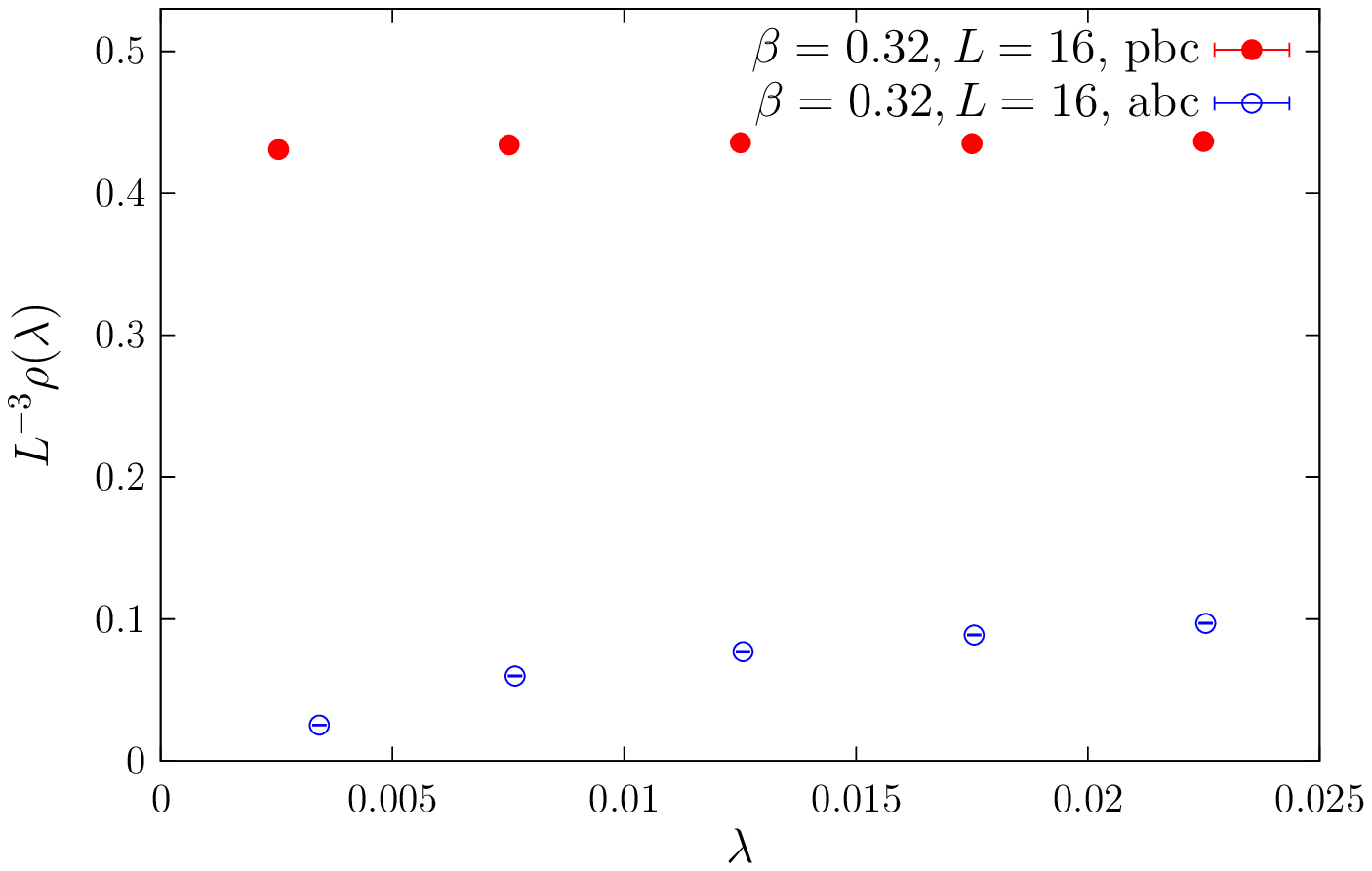} \hfil
  \includegraphics[width=0.49\textwidth]{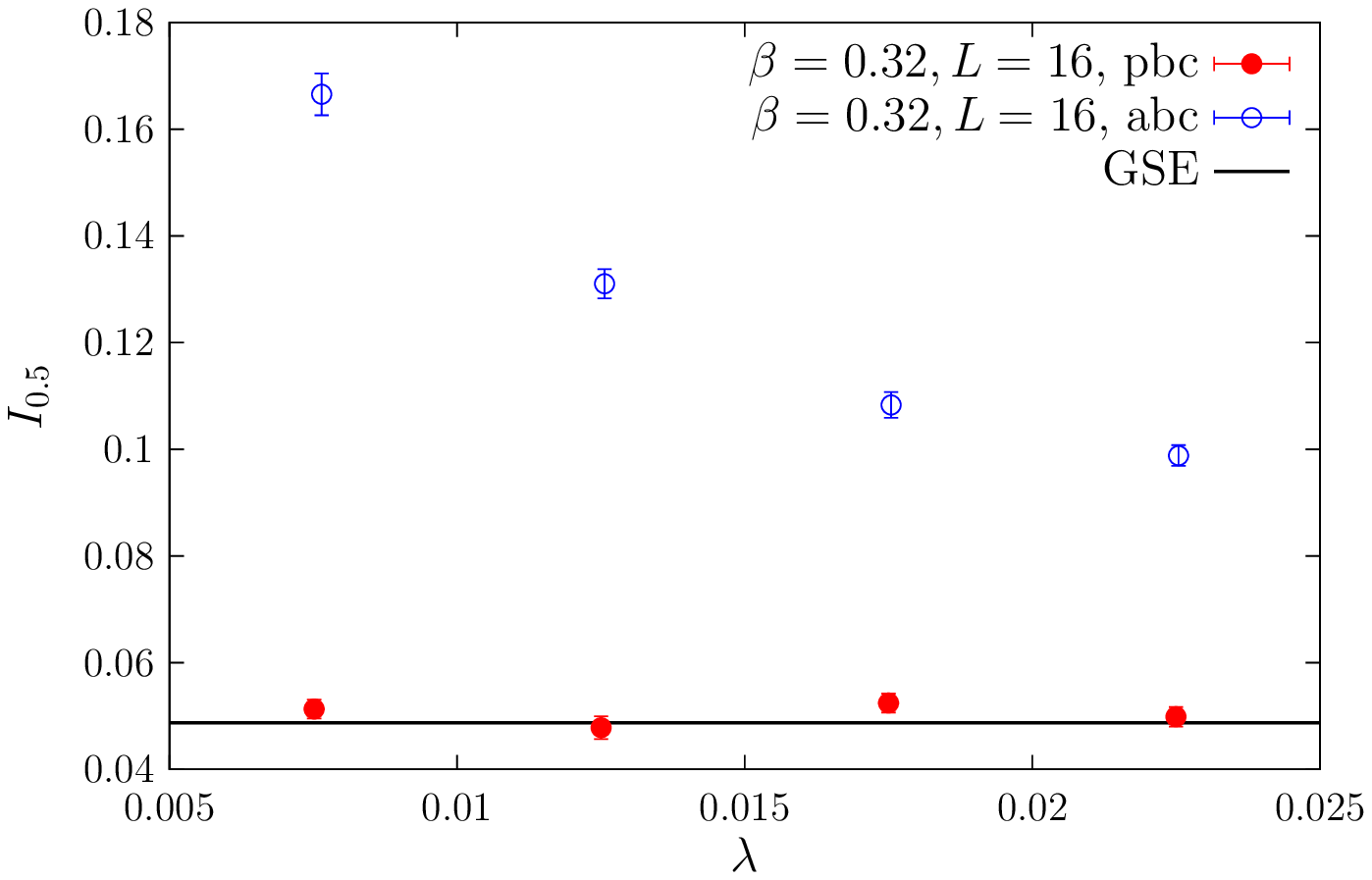}
  \caption{Comparison between the toy models obtained in
    correspondence to periodic (``pbc'', full points), or antiperiodic
    (``abc'', circles) boundary conditions for the fermions, in the
    ordered phase of the spin model: spectral density (left panel) and
    $I_{0.5}$ (right panel). Here $L=16$ and $\beta=0.32$.} 
  \label{fig:pbc}
\end{figure}

The latter case is naturally illustrated by changing the boundary
conditions for the fermions in the temporal direction from
antiperiodic to periodic. The derivation of the Dirac-Anderson
Hamiltonian proceeds exactly in the same way, and leads to the same
results, Eqs.~\eqref{eq:H_new_b} and \eqref{eq:H_new_b_elem}, up to
replacing the ``fermionic'' Matsubara frequencies $\omega_{ak}(\vec
x)$ with the ``bosonic'' frequencies $\omega^{\rm PBC}_{ak}(\vec x)$,
\begin{equation}
  \label{eq:pbc_matsu}
  \omega^{\rm PBC}_{ak}(\vec x) = {\textstyle\f{1}{N_T}}(
  \phi_a(\vec x)
  + 2\pi k)\,,\qquad k=0,1,\ldots,N_T-1\,.  
\end{equation}
The construction of the toy model is also unchanged, up to a similar
replacement in the diagonal terms, Eq.~\eqref{eq:toy_Ham_2}. In the
minimal setting $N_T=N_c=2$, this amounts to replace
$\cos\f{\phi_{\vec x}}{2} \to \sin\f{\phi_{\vec x}}{2}$. In the
ordered phase the distribution of unperturbed eigenvalues is now
peaked around zero, while the correlations between spatial links and 
thus the strength of the mixing of the temporal momentum components
are unchanged. Numerical results for this model are shown in
Fig.~\ref{fig:pbc}: ``chiral symmetry'' is broken and localisation
is absent. Notice that this is precisely what one expects, if the toy
model is to correctly reproduce the features of QCD~\cite{CC,Step},
which further supports the viability of our model to study the
qualitative behaviour of the Dirac spectrum.

\begin{figure}[t]
  \centering
  \includegraphics[width=0.49\textwidth]{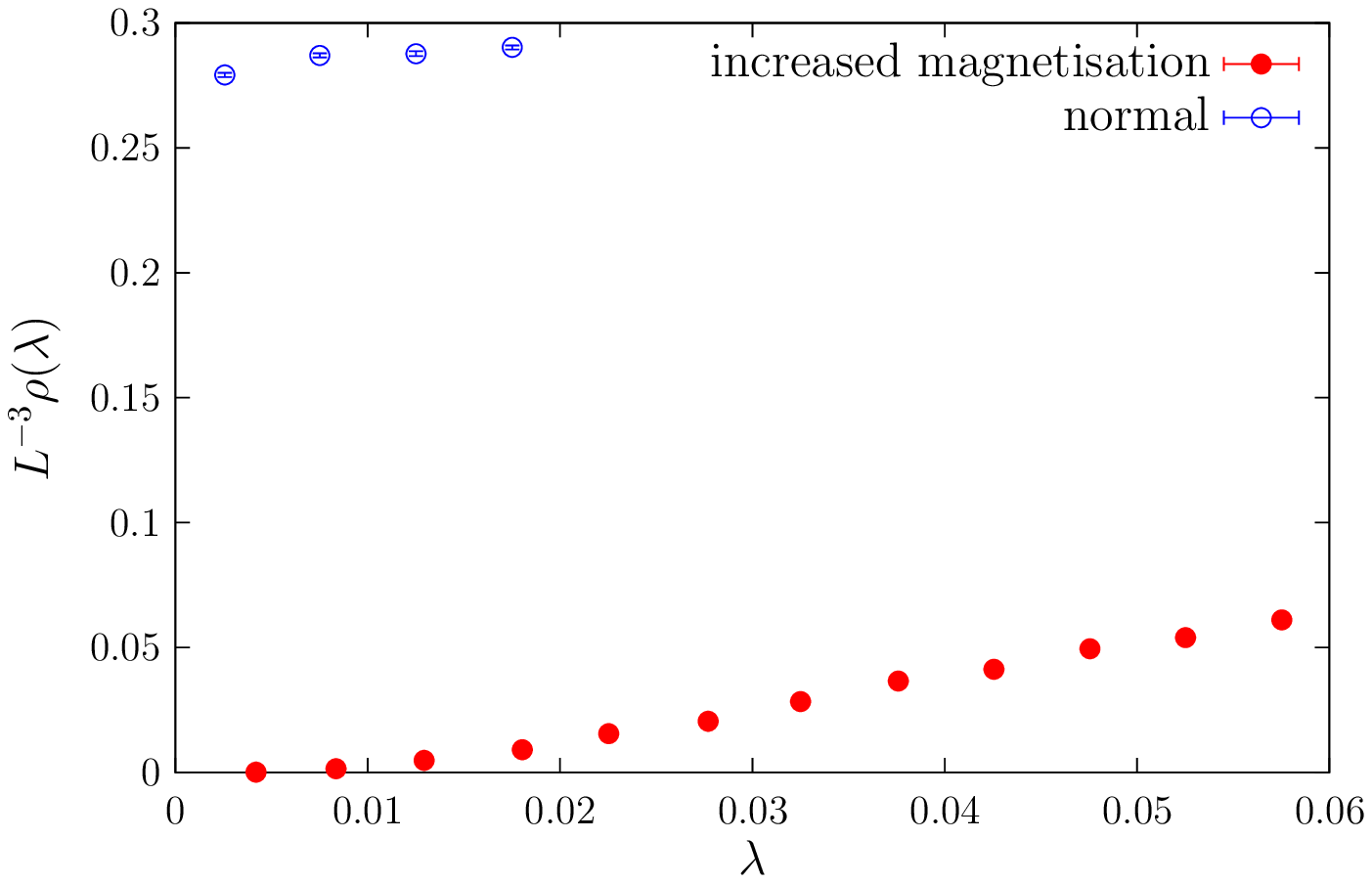} \hfil
  \includegraphics[width=0.49\textwidth]{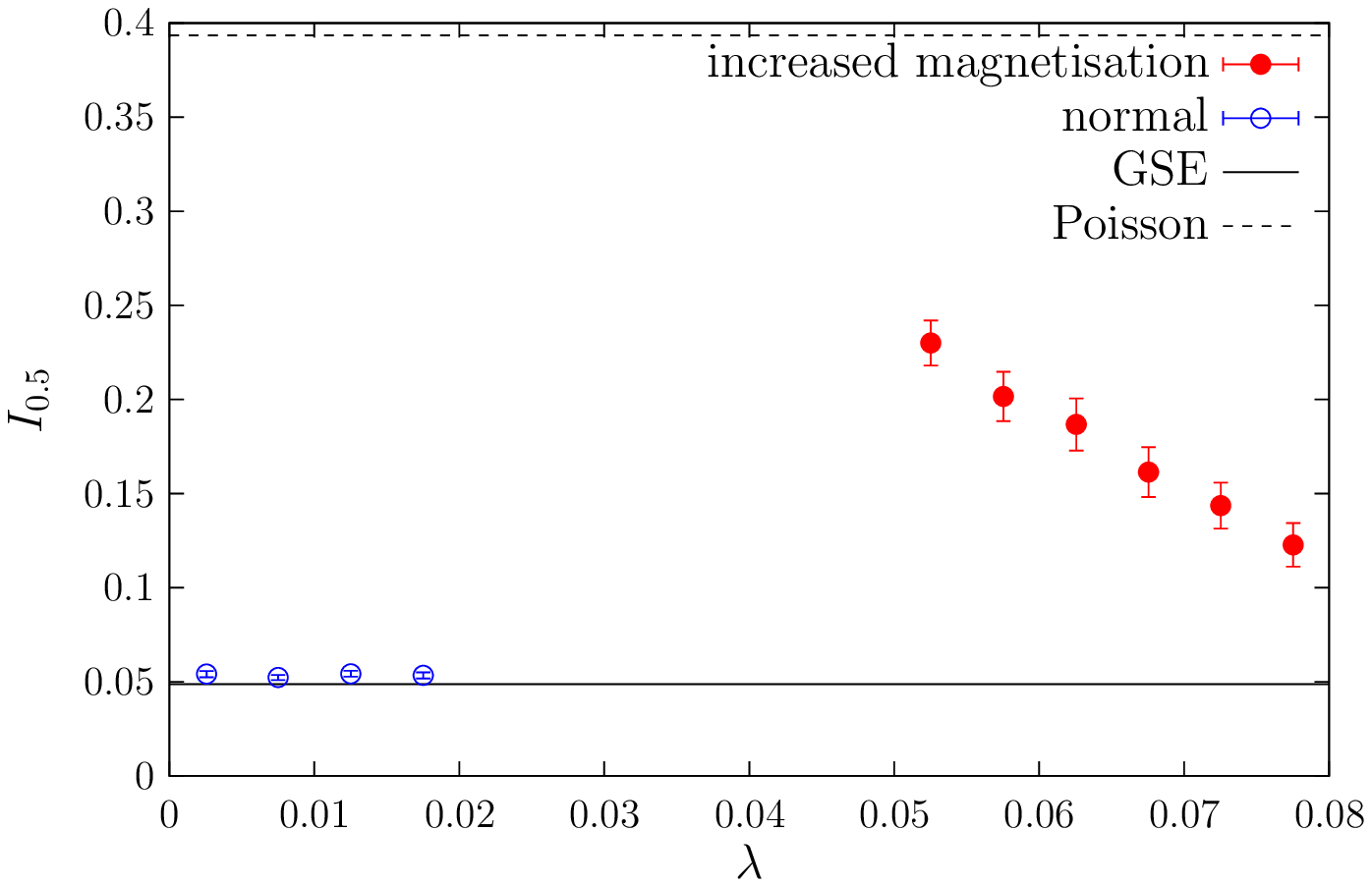}
  \caption{Comparison between the toy model in the disordered phase and
    the artificial model obtained by increasing the unperturbed
    eigenvalues via the mapping $f_\tau$ described in the text:
    spectral density (left panel) and $I_{0.5}$ (right panel). Here
    $L=16$, $\beta=0.29$ and $\tau=0.2$.} 
  \label{fig:incmag}
\end{figure}

To illustrate the former case we have employed a more artificial
construction, increasing ``by hand'' the ordering of the unperturbed
eigenvalues while keeping unchanged the hopping terms. This has been
achieved by mapping the absolute value of the unperturbed eigenvalues,
$z_{\vec x}=\cos\f{\phi_{\vec x}}{2}$, to $f_\tau(z_{\vec x}) =
\f{2z_{\vec x}^{\tau}}{1 + z_{\vec x}^{2\tau}}$ with $0<\tau<1$. As we
show in Fig.~\ref{fig:incmag}, 
in this way one can deplete the spectral region around zero: the
strength of the temporal-momentum-component mixing is not sufficient
to compensate for the lowered density of small unperturbed modes. 

\subsection{Discussion}
\label{sec:sub:disc}

Summarising our findings, the fate of ``chiral symmetry'' and of
localisation depends on the competition between ordering of the
unperturbed modes and mixing of temporal momentum components of the
wave functions. In the disordered phase, when there is no ordering of
the unperturbed modes and a sizeable amount of small such modes is
present, there is actually no competition and one expects chiral
symmetry to be broken. In the ordered phase, the reduction of mixing
can be compensated by changing the boundary conditions, so increasing
the amount of small unperturbed modes. Although we have checked here
only the case of periodic boundary conditions, one can in principle
change the effective boundary conditions in a continuous manner by
introducing an imaginary chemical potential, $\mu$: periodic boundary
conditions thus correspond to $|\mu N_T| = \pi$, and for $|\mu N_T|$
sufficiently close to $\pi$ we still expect to find a finite spectral
density at the origin. Sticking to $N_c=2$ and feeding this back into
the partition function by including the fermionic determinant, in the
ordered/deconfined phase one is lead to expect that if at $\mu=0$ the
trivial center sector is selected, at $|\mu N_T| = \pi$ the system is
again in the ordered/deconfined phase but in the other center
sector. For sufficiently high temperature one thus expects to find a
transition from one vacuum to the other when moving along the $\mu$
direction in the phase diagram, which then repeats due to periodicity
in $\mu$: these are nothing but the Roberge-Weiss
transitions~\cite{RW}. 

Our results shed some more light on the ``sea/islands'' mechanism in
QCD, and on its ineffectiveness at low temperature. As we have said
above, in the disordered phase the hopping term has typically sizeable
off-diagonal entries, which can effectively mix different
temporal-momentum components of the quark wave function. This is due
to the presence of extended spatial regions where the local
correlations across time slices are sufficiently weak, because of the
lack of order in the underlying spin system. As we saw above, this is
necessary for the accumulation of small eigenvalues, and one can think
of mixing as somehow ``pushing'' the eigenvalues towards zero. On top
of that, the lack of order also provides a sizeable density of small
unperturbed modes, which also favours such an accumulation. The
combination of these two effects thus leads naturally to a finite
spectral density near the origin, and to delocalisation of the low
modes. The ineffectiveness of the ``sea/islands'' mechanism at low
temperatures is thus due to the fact that in that case there simply
are no ``islands''. In contrast to this, in the ordered phase the
regions where mixing is effective are localised on the ``islands''
where the spins fluctuate away from the ordered value. In the ``sea''
of ordered spins the unperturbed eigenvalues are large (i.e., close to
1) and the mixing of different temporal-momentum components is
suppressed, so that the ``push'' towards the origin is weaker, and the
``sea'' does not contribute to the accumulation of eigenvalues around
the origin. Low modes thus originate from the ``islands'', and this
naturally leads to their localisation and to a small spectral density
near zero. 

There is another important aspect involved in chiral symmetry breaking
or restoration. For any finite volume, the spectral density decreases
to zero within a sufficiently small distance from the origin. What
actually determines the fate of chiral symmetry is how this distance
scales with the volume of the system. In the disordered phase, the
small eigenvalues originate from small unperturbed modes that occupy
a finite fraction of the volume. The ``push'' caused by mixing of 
the temporal-momentum components is thus expected to scale with the
system size, eventually leading to a finite spectral density at the
origin in the thermodynamic limit. In the ordered phase, the small
eigenvalues originate from small unperturbed modes localised on the
``islands'', and the strength of the ``push'' coming from mixing is
expected to correspond to the size of the ``islands'', which does not
scale with the system size, so that the spectral density at the origin
remains zero also in the thermodynamic limit.

Before concluding this Section, we want to list a few issues which
remain open. We have not checked whether the ``chiral'' transition in
our model is a genuine phase transition in the thermodynamic limit,
and if so whether it takes place exactly together with the transition
in the spin system. We also have not checked the dependence on the
strength $h$ of the $U(1)^{N_c-1}$-symmetry-breaking term in the
Hamiltonian of the spin system. As this would change the depth of the
symmetry-breaking potential, it can in principle affect the nature of
the transition in the spin system, and so in turn that of the
``chiral'' transition. Moreover, we have not investigated in details
how our results change when changing the parameter $\hat \beta$, which
affects the strength of the correlation between time slices. Since the
amount of small unperturbed modes is unaffected by such a change, it
is reasonable to expect that a larger $\hat\beta$, increasing the 
correlations and thus reducing the mixing between wave function
components corresponding to different temporal momenta in the ``sea''
region, will make the change in the spectral density when going over
to the ordered phase even more dramatic. Similarly, a smaller
$\hat\beta$ is expected to make this change less dramatic, and for
small enough $\hat\beta$ we expect that ``chiral symmetry'' is broken
also in the ordered phase of the spin model, by continuity with the
$\hat\beta=0$ result discussed above. However, it is not clear if the
value of $\hat\beta$ can affect the nature of the ``chiral''
transition, i.e., whether it is a true phase transition or a
crossover, and the value of $\beta$ at which it takes
place. Nevertheless, we think that the connection between correlation
of spatial links across time slices and fate of ``chiral symmetry'' is
strongly supported by our findings.

\section{Conclusions}
\label{sec:concl}

In this paper we have studied the problems of chiral symmetry breaking
and localisation in finite-temperature QCD by looking at the lattice
Dirac operator as a random Hamiltonian. We have explicitly recast the
staggered Dirac operator at finite temperature in the form of a
non-conventional 3D Anderson Hamiltonian (``Dirac-Anderson
Hamiltonian''), which describes fermions carrying colour and an extra
internal degree of freedom, corresponding to the lattice temporal
momenta. The on-site noise is provided by the phases of the Polyakov
lines, and their ordering, or the lack thereof, reflects in the
distribution of the diagonal entries of the Hamiltonian. The Polyakov
lines affect the hopping terms as well. Indeed, in the deconfined
phase they induce strong correlations among spatial links on different
time slices, in the region where they are aligned with the
identity. This in turn weakens the coupling among the components of
the wave function corresponding to different lattice temporal momenta
on neighbouring sites. In the confined phase, on the other hand, such
strong correlations are absent due to the absence of order, and these
wave function components can mix effectively. We think that this
difference in the hopping terms is essential to explain the
accumulation or not of eigenvalues near the origin, and ultimately for
the spontaneous breaking or not of chiral symmetry. The other
important difference between the two phases concerns the density and
spatial distribution of small unperturbed eigenvalues, i.e., of small
diagonal entries, which is also caused by the different ordering
properties of the Polyakov lines. We think that these two properties 
of the small unperturbed eigenvalues are also essential in explaining
the fate of chiral symmetry. Furthermore, the fact that the small
unperturbed eigenvalues appear in localised spatial regions at high
temperature leads to the localisation of the low Dirac eigenmodes. 
This suggests that the confinement/deconfinement transition triggers
both the chiral transition and the localisation of the low modes. 

To test this picture we have constructed a toy model, made up of a
spin system with dynamics similar to that of the Polyakov line phases
in QCD; of unitary matrices obeying dynamics analogous to the dynamics
of spatial gauge links in the background of fixed Polyakov lines; 
and of a random Hamiltonian with the same structure as the
Dirac-Anderson Hamiltonian discussed above, with on-site noise
provided by the spins. This toy model is designed to keep precisely
the features of QCD which we believe to be relevant to the phenomena
of chiral symmetry breaking (which means here a nonzero spectral
density at the origin) and localisation. A numerical study of the toy
model, in the simplest case of two colours and two time slices, shows
that it indeed displays both these phenomena, with the same
qualitative dependence on the ordering of the noise source
(spins/Polyakov lines) as in QCD. When the noise source is disordered
there is chiral symmetry breaking but no localisation; when the noise
source is ordered chiral symmetry is restored and low modes are
localised, up to a point in the spectrum (``mobility edge'') which is
pushed towards larger values as the ordering is increased. If, on the
other hand, one artificially removes the correlation between the
time slices in the ordered phase, then chiral symmetry remains broken
and there is no localisation. Moreover, if the mixing between
components corresponding to different time-momenta is artificially
removed in the disordered phase, then chiral symmetry is restored and
the lowest modes localise. These findings support, and actually
suggested, that the correlation among time slices and the related
mixing of temporal-momentum components of the quark wave functions
play an essential role in the chiral transition and in the appearance
of localised modes. The importance of the role played by the small
unperturbed eigenvalues is made evident by the results obtained when
one imposes on the fermions periodic rather than antiperiodic
boundary conditions in the temporal direction. In this case the
properly modified toy model again reproduces qualitatively the QCD
results, with accumulation of eigenvalues around the origin and no
localisation also in the ordered phase. Since the hopping terms are
exactly the same as with antiperiodic boundary conditions, one is led
to conclude that even with weak mixing between temporal-momentum
components one can achieve chiral symmetry breaking, if the density of
small unperturbed modes is large enough. Conversely, by artificially
increasing the unperturbed modes without touching the hopping terms,
one can restore chiral symmetry in the disordered phase and make the
low Dirac modes localised, which indicates that strong mixing may be
insufficient to accumulate eigenvalues around zero if the density of
small unperturbed modes is too low. 

The results obtained in the toy model support our expectation that in
QCD the fate of chiral symmetry and of localisation are closely
related, and furthermore that both depend on the amount of small
unperturbed modes and on the mixing of temporal-momentum components of
the quark wave functions, therefore ultimately on the distribution of
the phases of the Polyakov lines. 

The picture discussed here makes no direct reference to topology. As
is well known, in the ``topological'' explanation of chiral symmetry
breaking the finite density of near-zero modes originates from
fermionic zero modes supported by topological objects, which broaden
into a band due to mixing. The localised nature of these modes would
also explain localisation at high temperature. In the Dirac-Anderson
picture, the ``unperturbed'' small modes have a different origin,
being the eigenmodes of the temporal part of the Dirac operator, and
moreover the way they mix (i.e., the nature of the hopping terms) is
also expected to play an important role in the accumulation of
near-zero modes and in their localisation properties. It is of course 
well possible that the two pictures are just complementary point of
views on the same phenomenon, corresponding to a different way to
separate the full Dirac operator into  a ``free'' and an
``interaction'' part. In light of the close connection between 
the chiral transition and localisation, and of the central role played
by the Polyakov lines in both phenomena, we expect that if this is the
case, then the topological objects relevant to chiral symmetry
breaking at low temperature, and to localisation at high temperature, 
would also be relevant to the deconfinement transition. Indeed, there
are 
numerical results pointing to a close relation between
localisation and certain topological objects which are expected to
play a role in the deconfinement transition~\cite{Cossu:2016scb}. This
issue certainly deserves more work. Attention should also be paid to 
the possible relation between localisation and chiral symmetry 
restoration, on one side, and ``non-topological'' approaches to
confinement like, e.g.,
``fluxons''~\cite{Mandelstam:1974pi,'tHooft:1977hy}. 

It would be interesting to further investigate in our toy model the
behaviour of the spectrum in the vicinity of the phase transition in
the spin model. This would clarify if the ``chiral transition'' seen
there is actually a genuine phase transition, and how close it takes
place to the magnetic transition. This could provide useful insight in
the critical properties at the chiral transition of the ``parent''
physical system, namely, QCD. Furthermore, one could check how the
transition is affected by the strength of the coupling between spatial
links on different time slices, and by the depth of the minimum of the
spin potential, which are parameters of the model besides the
temperature of the spin system. 

An obvious extension of this work would be to check the ideas
presented above directly with QCD gauge configurations, using the
Dirac-Anderson form of the Dirac operator to tweak the hopping terms
independently of the underlying Polyakov-line dynamics. While the toy
model studied here is quenched, with no backreaction of the
quark eigenvalues in the partition function, the main ideas are
expected to apply in the presence of dynamical fermions as well.

It would be interesting to try to apply the ideas discussed in this
paper in the case when a constant (Abelian) magnetic field is turned
on. This could shed some light on the issue of (inverse) magnetic
catalysis of the quark condensate~\cite{imc1,imc2}. Another
interesting application would be to the case of nonzero imaginary
chemical potential, already very briefly discussed here.

Another interesting testing ground for the proposed mechanism is the
explanation of the separate occurrence of the deconfinement and chiral
transitions in $SU(3)$ gauge theory with adjoint
fermions~\cite{adjoint}. Since the derivation of the Dirac-Anderson
Hamiltonian did not use in an essential way that we were considering
fundamental fermions, the same form holds for adjoint fermions,
replacing gauge links with their adjoint counterpart, and the $N_c$
phases of the fundamental Polyakov line with the $N_c^2-1$ phases of
the adjoint Polyakov line. 

In conclusion, we believe that the ``Dirac-Anderson'' approach of the
present paper to the study of the quark eigenvalues and eigenfunctions
can lead to a better understanding of the phase structure of QCD and
related theories.

\section*{Acknowledgements}
FP is supported by OTKA under the grant OTKA-K-113034, and partially
under the grant OTKA-NF-104034. TGK and (partially) MG are supported
by the Hungarian Academy of Sciences under ``Lend\"ulet'' grant
No. LP2011-011.  MG wants to thank Guido Cossu for useful discussions 
and for sharing his preliminary results. MG also wants to thank the
Institute for Nuclear Research in Debrecen, where most of the work in
this paper has been carried out.

\appendix

\section{Properties of the hopping term}
\label{sec:app}

In this Appendix we describe in some detail the properties of the QCD
Dirac-Anderson ``Hamiltonian'', and in particular of its hopping
terms, ${\cal V}_{\pm j}$. First of all, notice that  
\begin{equation}
  \label{eq:app_VVdag}
  \begin{aligned}
    \left[{\cal V}^\dag_{\pm j}(\vec x,\vec y)\right]_{ak,bl} &=
\left[{\cal V}^*_{\pm j}(\vec y,\vec x)\right]_{bl,ak} =
    \delta_{\vec x \mp \hat\jmath,\vec y}
    \f{1}{N_T} \sum_{t=0}^{N_T-1}
    e^{-i\f{2\pi t}{N_T}(k-l)} \left[\tilU^*_{\pm j}(t,\vec
      y)\right]_{ba}\\ 
& =  \delta_{\vec x \mp \hat\jmath,\vec y}
    \f{1}{N_T} \sum_{t=0}^{N_T-1}
    e^{i\f{2\pi t}{N_T}(l-k)} \left[\tilU^\dag_{\pm j}(t,
      \vec x \mp  \hat\jmath)\right]_{ab}
\\ &=  \delta_{\vec x \mp \hat\jmath,\vec y}
    \f{1}{N_T} \sum_{t=0}^{N_T-1}
    e^{i\f{2\pi t}{N_T}(l-k)} \left[\tilU_{\mp j}(t,\vec x )\right]_{ab} =
\left[{\cal V}_{\mp j}(\vec x,\vec y)\right]_{ak,bl}\,,
    \end{aligned}
\end{equation}
as it should. From now on we will often omit matrix indices, so we
remind the reader that $\tilU_{\pm j}(t,\vec x)$ has only colour indices,
while $D(\vec x)$ and $V_{\pm j}(\vec x)$ have both colour and
temporal-momentum indices. The identity in these spaces will be
denoted by $\mathbf{1}_{\rm c}$ and $\mathbf{1}_{\rm tm}$,
respectively. Since $\tilU_{\pm j}(t,\vec x)$ are unitary matrices,
\begin{equation}
  \label{eq:app_prop_U}
  \begin{aligned}
    &  \tilU_{\pm j}(t,\vec x) [\tilU_{\pm j}(t,\vec x)]^\dag =
    \mathbf{1}_{\rm c}\,, \\
    & \det \tilU_{\pm j}(t,\vec x) =
    e^{i\f{2\pi t}{N_T}[q(\vec x\pm \hat
      \jmath)-q(\vec x)]} \,, \quad 2\pi q(\vec x) =
    \sum_{a=1}^{N_c} \phi_{a}(\vec x)\,, \quad q(\vec x)\in
    \mathbb{Z}\,, 
  \end{aligned}
\end{equation}
and since $V_{\pm j}(\vec x)$ is the Fourier transform with respect to
time of the unitary matrix $\bar{U}_{\pm j}(\vec x)$, 
\begin{equation}
  \label{eq:app_prop_V}
    V_{\pm j}(\vec x) = 
    F\bar{U}_{\pm j}(\vec x) F^\dag\,, 
\end{equation}
where
\begin{equation}
  \label{eq:app_prop_V_bis}
 \left[\bar{U}_{\pm j}(\vec x)\right]_{at,bt'} = 
\delta_{tt'}\left[\tilU_{\pm j}(t,\vec x)\right]_{ab} \,,
\qquad     F_{kt}= \f{1}{\sqrt{N_T}}\,e^{-i\f{2\pi t}{N_T}kt}\,,
\end{equation}
with $FF^\dag =\mathbf{1}_{\rm tm}$, we conclude that $V_{\pm j}(\vec
x)$ is also unitary in colour and temporal-momentum space,
\begin{equation}
  \label{eq:app_prop_V1}
  V_{\pm j}(\vec x)V_{\pm j}(\vec x)^\dag = \mathbf{1}_{\rm
    c}\mathbf{1}_{\rm tm}\,.
\end{equation}
Moreover, since (recall that $N_T$ is even)
\begin{equation}
  \label{eq:app_prop_V2}
\det V_{\pm j}(\vec x) = \prod_{t=0}^{N_T-1}
e^{i\f{2\pi t}{N_T}[q(\vec x\pm \hat
      \jmath)-q(\vec x)]} = e^{i\pi [q(\vec x\pm \hat
      \jmath)-q(\vec x)](N_T-1)} = e^{-i\pi [q(\vec x\pm \hat
      \jmath)-q(\vec x)]} = \pm 1\,,
\end{equation}
we have that $V_{\pm j}(\vec x)$ is also unimodular up to a
sign. If we choose one and the same convention for the phases of the
local Polyakov lines, i.e., we fix $q(\vec x)=q ~\forall \vec x$,
then $V_{\pm j}(\vec x)\in SU(N_c\times N_T), ~\forall \vec
x,j$. However, one can choose different phase conventions at 
different spatial points, and still obtain the same physical
results. We will return below to this issue. We observe also the
following cyclicity property of $V_{\pm j}(\vec x)$,
\begin{equation}
  \label{eq:app_prop_V3}
  \left[V_{\pm j}(\vec x)\right]_{a(k+n)_{N_T},b(l+n)_{N_T}} =
  \left[V_{\pm j}(\vec x)\right]_{ak,bl} \,,\,\, \forall
  n\in\mathbb{Z}\,. 
\end{equation}
One can easily verify that this property is preserved under
multiplications, so $V_{\pm j}(\vec x)$ belong to the ``$(N_T\times
N_T)$-block cyclic'' subgroup of $SU(N_c\times N_T)$. 

Let us now return to the issue of the choice of phase conventions. All 
$\phi_a(\vec x)$ are defined modulo $2\pi$, so after a redefinition
$\phi_a(\vec x)\to \phi_a(\vec x) + 2\pi q_a(\vec x)$ with
$q_a(\vec x)\in \mathbb{Z}$ one should obtain equivalent results. We
will denote quantities after the redefinition with the superscript
$\{q\}$. We have for the effective Matsubara frequencies
\begin{equation}
  \label{eq:app_phases_conv}
  \exp[i\omega_{ak}^{\{q\}}(\vec x)] =     
  \exp\left[i\textstyle{\f{1}{N_T}}(
    \pi + \phi_a(\vec x) + 2\pi q_a(\vec x) + 2\pi k)\right]
  = \exp[i\omega_{a(k+q_a(\vec x))_{N_T}}(\vec x)] \,,
\end{equation}
so that the temporal-momentum indices of $D(\vec x)$ and $V_{\pm
  j}(\vec x)$ undergo a colour-dependent cyclic permutation $k \to
k+q_a(\vec x)\mod N_T$. This can be written formally using the
permutation matrices $\Pi^{(n)}$, 
\begin{equation}
  \label{eq:app_Z_perm}
  \Pi^{(n)}_{kk'} = \delta_{(k+n)_{N_T} k'}\,,\quad
  \Pi^{(n)}[\Pi^{(n)}]^\dag = \Pi^{(n)}[\Pi^{(n)}]^T = \mathbf{1}_{\rm tm}\,,
\end{equation}
by setting
\begin{equation}
  \label{eq:app_Z_perm_full}
  \left[Z^{\{q\}}(\vec x)\right]_{ak,bl}=
  \Pi^{(q_a(\vec x))}_{kl}\delta_{ab}\,.
\end{equation}
We then have
\begin{equation}
  \label{eq:app_phases_conv2}
  D^{\{q\}}(\vec x) =
  {Z}^{\{q\}}(\vec x)D(\vec x){Z}^{\{q\}}(\vec x)^\dag\,, 
  \qquad   V^{\{q\}}_{\pm j}(\vec x) =
  {Z}^{\{q\}}(\vec x)  V_{\pm j}(\vec x)
  {Z}^{\{q\}}(\vec x\pm\hat\jmath)^\dag\,.
\end{equation}
Notice that $V_{\pm j}(x)$ is left invariant if $q_a(\vec x)=
q_b(\vec x\pm \hat \jmath)~\forall a,b$, as a consequence of its
cyclicity property. Finally, defining
\begin{equation}
  \label{eq:app_Z_perm_full2}
  \left[{\cal Z}^{\{q\}}(\vec x,\vec y)\right]_{ak,bl}=
 [ Z^{\{q\}}(\vec x)]_{kl}\,\delta_{ab}\,\delta_{\vec x\vec y}\,,\qquad
 {\cal Z}^{\{q\}}{\cal Z}^{\{q\}}{}^\dag = \mathbf{1}_{\rm
   c}\mathbf{1}_{\rm tm}\mathbf{1}_{\rm sp}\,,
\end{equation}
with $\mathbf{1}_{\rm sp}$ the identity in spatial-coordinate space,
we can write in compact form
\begin{equation}
  \label{eq:app_phases_conv4}
  {\cal H}^{\{q\}} = {\cal Z}^{\{q\}}  {\cal H}
{\cal Z}^{\{q\}\,\dag}\,,
\end{equation}
which amounts to say that a redefinition of the phases corresponds to
a unitary transformation of the Hamiltonian, which therefore leaves the
spectrum unchanged.

It is interesting to notice that if we make the unitary transformation
Eq.~\eqref{eq:app_phases_conv4} setting $q_a(\vec x) =
\f{N_T}{2}~\forall \vec x,a$, then, denoting $\tilde {\cal Z}= {\cal
  Z}^{\{q_a(\vec x)=\f{N_T}{2}\}}$, we find [see Eq.~\eqref{eq:7bis}]
\begin{equation}
  \label{eq:app_phases_conv5}
  \tilde{\cal Z}  {\cal H}  \tilde{\cal Z}^\dag
  =    - {\cal D}(\vec x,\vec y) +
  \sum_{j=1}^3\f{\eta_j(\vec x)}{2i}\left\{{\cal V}_{+j}(\vec x,\vec
    y)-{\cal V}_{-j}(\vec x,\vec y)\right\}
  = - \eta_4{\cal H}\eta_4
  \,,
\end{equation}
where $(\eta_4{\cal H}\eta_4)(\vec x,\vec y)=\eta_4(\vec x){\cal
  H}(\vec x,\vec y)\eta_4(\vec y)$. We conclude that ${\cal
  Q}\equiv\eta_4 \tilde{\cal Z}$ satisfies  $\{{\cal Q},{\cal H}\}=0$,
which implies that the spectrum is symmetric with respect to
$\lambda=0$, as it should be for staggered fermions.  

One final remark is in order concerning the case of gauge group
$SU(2)$. In this case one has $\phi_1 = -\phi_2 \equiv \phi$.
It is known that in this case the Dirac operator has an antiunitary
symmetry ${\cal T}$ with ${\cal T}^2=-1$~\cite{VWrev}. In the new basis,
taking the complex conjugate of ${\cal H}$ has the effect of (i)
exchanging the indices $k,l$ in temporal-momentum space, (ii) changing
the sign of the phases $\phi$ appearing in $\tilU$
[Eq.~\eqref{eq:H_new_b_elem}] and taking the complex conjugate of the
$SU(2)$ matrices $\tdU_{\pm j}$, and (iii) changing the overall sign
of the hopping terms. Point (ii) can be ``undone'' by taking the
matrix conjugate with respect to $\sigma_2$ in colour space (this
remains true also in the presence of nontrivial phases $\phi$, as can
be directly checked); this also leads to the diagonal element $k$
being switched with the element $\f{N_T}{2}-1-k$. Indeed, matrix
conjugation by $\sigma_2$ exchanges the diagonal terms corresponding
to $\phi$ and $-\phi$, and this is equivalent to switching
temporal-momentum components, since 
\begin{equation}
  \label{eq:app_T_SU2}
  \sin\tf{\pi - \phi + 2\pi k}{N_T} =
  \sin\tf{\pi + \phi + 2\pi \left(\tf{N_T}{2}-1-k\right)}{N_T}\,.
\end{equation}
This corresponds to a permutation $\Pi$ of the temporal-momentum
components defined so that $k\to \f{N_T}{2}-1-k \mod N_T$ (notice
$\Pi^2=1$). Since the hopping term depends on $k,l$ only through
$k-l$, one has $l-k = \left(\f{N_T}{2}-1-k\right) -
\left(\f{N_T}{2}-1-l\right)$, and so by applying $\Pi$ we undo both
point (i) and the above-mentioned effect on the diagonal
term. Finally, taking the matrix conjugate with respect to $\eta_4$ in
spatial-coordinate space we undo point (iii). All in all, ${\cal
  T}=\eta_4\Pi\sigma_2 K$, with $K$ the complex conjugation, is an
antiunitary symmetry of the Hamiltonian with ${\cal
  T}^2=-\sigma_2^2=-1$.


\begin{thebibliography}{99}

\bibitem{BC} T.~Banks and A.~Casher, Nucl. Phys. B 
\nyp{169}{103}{1980}.

\bibitem{VWrev}
  J.~J.~M.~Verbaarschot and T.~Wettig,
  Ann.\ Rev.\ Nucl.\ Part.\ Sci.\  
\nyp{50}{343}{2000} 
  [hep-ph/0003017].

\bibitem{deF}
  P.~de Forcrand,
  AIP Conf.\ Proc.\  
\nyp{892}{29}{2007} 
  [hep-lat/0611034].

\bibitem{GGO} 
  A.~M.~Garc\'ia-Garc\'ia and J.~C.~Osborn,
  Nucl.\ Phys.\ A 
\nyp{770}{141}{2006} 
  [hep-lat/0512025].

\bibitem{GGO2}
  A.~M.~Garc\'ia-Garc\'ia and J.~C.~Osborn,
  Phys.\ Rev.\  D 
\nyp{75}{034503}{2007}
  [hep-lat/0611019].

\bibitem{KGT} T.~G.~Kov\'acs,
  Phys.\ Rev.\ Lett.\ 
\nyp{104}{031601}{2010}   
[arXiv:0906.5373 [hep-lat]]. 


\bibitem{KP} T.~G.~Kov\'acs and
    F.~Pittler, Phys.\ Rev.\ Lett.\ 
\nyp{105}{192001}{2010} 
  [arXiv:1006.1205 [hep-lat]].

\bibitem{BKS} 
  F.~Bruckmann, T.~G.~Kov\'acs and S.~Schierenberg,
  Phys.\ Rev.\ D 
\nyp{84}{034505}{2011} 
 [arXiv:1105.5336 [hep-lat]].


\bibitem{KP2} T.~G.~Kov\'acs and
    F.~Pittler, Phys.\ Rev.\ D 
\nyp{86}{114515}{2012}  
  [arXiv:1208.3475 [hep-lat]].

\bibitem{feri} M.~Giordano, T.~G.~Kov\'acs and
    F.~Pittler,  PoS  LATTICE 
\nyp{2013}{212}{2013}
     [arXiv:1311.1770    [hep-lat]].  



\bibitem{crit} M.~Giordano, T.~G.~Kov\'acs and F.~Pittler, 
    Phys.\ Rev.\ Lett.\ 
\nyp{112}{102002}{2014} 
  [arXiv:1312.1179 [hep-lat]].


\bibitem{Cossu:2014aua}
G.~Cossu, H.~Fukaya, S.~ Hashimoto, T.~Kaneko, J.~Noaki and A.~Tomiya
[JLQCD Collaboration], 
  PoS LATTICE \nyp{2014}{210}{2015}
  [arXiv:1412.5703 [hep-lat]].

\bibitem{Cossu:2016scb}
  G.~Cossu and S.~Hashimoto,
  arXiv:1604.00768 [hep-lat].


\bibitem{Aoki:2005vt} 
  Y.~Aoki, Z.~Fodor, S.~D.~Katz and K.~K.~Szab\'o,
  \JHEP \nyp{01}{089}{2006} 
  [hep-lat/0510084].

\bibitem{Borsanyi:2010cj} 
  S.~Bors\'anyi, G.~Endr\H odi, Z.~Fodor, A.~Jakov\'ac, S.~D.~Katz, S.~Krieg,
  C.~Ratti and K.~K.~Szab\'o,
  \JHEP \nyp{11}{077}{2010} 
  [arXiv:1007.2580 [hep-lat]].

\bibitem{Giordano:2014qna}
  M.~Giordano, T.~G.~Kov\'acs and F.~Pittler,
   Int.\ J.\ Mod.\ Phys.\ A \nyp{29}{1445005}{2014}. 
  [arXiv:1409.5210 [hep-lat]].

\bibitem{unimproved}
P.~de Forcrand and O.~Philipsen, \JHEP 
\nyp{11}{012}{2008}
[arXiv:0808.1096 [hep-lat]].

\bibitem{GKKP} M.~Giordano, T.~G.~Kov\'acs, S.~D.~Katz and F.~Pittler, 
PoS LATTICE  \nyp{2014}{214}{2014} [arXiv:1410.8392 [hep-lat]].

\bibitem{Anderson58}
  P.~W.~Anderson, Phys.\ Rev.\ 
\nyp{109}{1492}{1958}.

\bibitem{LR}
  P.~A.~Lee and T.~V.~Ramakrishnan,
  Rev.\ Mod.\ Phys.\  
\nyp{57}{287}{1985}.

\bibitem{EM}
  F.~Evers and A.~D.~Mirlin,
  Rev.\ Mod.\ Phys.\  
\nyp{80}{1355}{2008}
[arXiv:0707.4378 [cond-mat.mes-hall]]. 


\bibitem{nu_unitary}
K.~Slevin and T.~Ohtsuki, Phys.\ Rev.\ Lett.\ 
\nyp{78}{4083}{1997}
[cond-mat/9704192 [cond-mat.dis-nn]].

\bibitem{Mehta}
  M.~Mehta, {\it Random Matrices}
(Academic Press, San Diego, 1991).


\bibitem{UGPKV} 
  L.~Ujfalusi, M.~Giordano, F.~Pittler, T.~G.~Kov\'acs and I.~Varga,
Phys.\ Rev.\ D \nyp{92}{094513}{2015} 
[arXiv:1507.02162 [cond-mat.dis-nn]].

\bibitem{UV}
L.~Ujfalusi and I.~Varga,
  Phys.\ Rev.\ B \nyp{91}{184206}{2015}
  [arXiv:1501.02147 [cond-mat.dis-nn]].




\bibitem{offdiag} E.~N.~Economou and P.~D.~Antoniou, Solid State Commun.\
\nyp{21}{285}{1977}.

\bibitem{offdiag2}
D.~Weaire and V.~Srivastava, Solid State Commun.\
\nyp{23}{863}{1977}.

\bibitem{GGC}   A.~M.~Garc\'ia-Garc\'ia and E.~Cuevas,
Phys.\ Rev.\ B 
\nyp{74}{113101}{2006}
[cond-mat/0602331 [cond-mat.dis-nn]].


\bibitem{GKP}
  M.~Giordano, T.~G.~Kov\'acs and F.~Pittler,
   \JHEP \nyp{04}{112}{2015}
  [arXiv:1502.02532 [hep-lat]].





\bibitem{Yaffe:1982qf}
  L.~G.~Yaffe and B.~Svetitsky,
  Phys.\ Rev.\ D 
\nyp{26}{963}{1982}.

\bibitem{DeGrand:1983fk}
  T.~A.~DeGrand and C.~E.~DeTar,
  Nucl.\ Phys.\ B 
\nyp{225}{590}{1983}.

\bibitem{SSSLS} 
    B.~I.~Shklovskii, B.~Shapiro,  
  B.~R.~Sears, P.~Lambrianides and H.~B.~Shore, Phys. Rev. B {\bf
    47} (1993) 11487.

\bibitem{HS} E.~Hofstetter
    and M.~Schreiber, Phys. Rev. B {\bf 49} (1994) 14726.


\bibitem{CC} S.~Chandrasekharan and N.~H.~Christ, Nucl.\ Phys.\ Proc.\
  Suppl.\ {\bf 47} (1996) 527 [arXiv:hep-lat/9509095].

\bibitem{Step} M.~A.~Stephanov, Phys.\ Lett.\ B {\bf 375} (1996) 249
  [arXiv:hep-lat/9601001]. 


\bibitem{RW}
  A.~Roberge and N.~Weiss,
  Nucl.\ Phys.\ B 
\nyp{275}{734}{1986}.



\bibitem{Mandelstam:1974pi}
  S.~Mandelstam,
  Phys.\ Rept.\  {\bf 23} (1976) 245.

\bibitem{'tHooft:1977hy}
  G.~'t Hooft,
  Nucl.\ Phys.\ B {\bf 138} (1978) 1.

\bibitem{imc1}
  G.~S.~Bali, F.~Bruckmann, G.~Endr\H odi, Z.~Fodor, S.~D.~Katz,
  S.~Krieg, A.~Sch\"afer and K.~K.~Szab\'o, 
  \JHEP \nyp{02}{044}{2012}
  [arXiv:1111.4956 [hep-lat]].
\bibitem{imc2}
  F.~Bruckmann, G.~Endr\H odi and T.~G.~Kov\'acs,
  \JHEP \nyp{04}{112}{2013}
  [arXiv:1303.3972 [hep-lat]].

\bibitem{adjoint}
  F.~Karsch and M.~L\"utgemeier,
  Nucl.\ Phys.\ B 
\nyp{550}{449}{1999}
  [hep-lat/9812023].













\end{thebibliography}
\end{document}